\documentclass[twoside,twocolumn,10pt]{extarticle}

\usepackage[numbers,sort&compress,comma]{natbib}
\usepackage[version=3]{mhchem}
\usepackage[left=1.5cm, right=1.5cm, top=1.785cm, bottom=2.0cm]{geometry}
\usepackage{balance}
\usepackage{mathptmx}
\usepackage{amsmath}
\usepackage{booktabs}
\usepackage{siunitx}
\usepackage{placeins}
\usepackage{graphicx}

\usepackage[utf8]{inputenc}
\usepackage{textgreek}
\usepackage[format=plain,justification=justified,singlelinecheck=false,font={stretch=1.125,small},labelfont=bf,labelsep=period]{caption}
\usepackage{subcaption}
\usepackage{lastpage}
\usepackage{float}
\usepackage{fancyhdr}
\usepackage{fnpos}
\usepackage[english]{babel}
\addto{\captionsenglish}{%
  
}
\usepackage{array}
\usepackage[T1]{fontenc}
\usepackage[dvipsnames]{xcolor}
\usepackage{setspace}
\usepackage[compact]{titlesec}
\usepackage[hidelinks]{hyperref}
\usepackage{epstopdf}

\definecolor{cream}{RGB}{222,217,201}
\renewcommand*\rmdefault{ptm}

\begin{document}
\raggedbottom
\setcitestyle{numbers,square}
\pagestyle{fancy}
\thispagestyle{plain}
\fancypagestyle{plain}{
%%%HEADER%%%
\renewcommand{\headrulewidth}{0pt}
}
%%%END OF HEADER%%%

%%%PAGE SETUP%%%
\makeFNbottom
\makeatletter
\renewcommand\LARGE{\@setfontsize\LARGE{14pt}{16}}
\renewcommand\Large{\@setfontsize\Large{12pt}{14}}
\renewcommand\large{\@setfontsize\large{10pt}{12}}
\renewcommand\footnotesize{\@setfontsize\footnotesize{7pt}{10}}
\makeatother

\renewcommand{\thefootnote}{\fnsymbol{footnote}}
\renewcommand\footnoterule{\vspace*{1pt}%
\color{cream}\hrule width 3.5in height 0.4pt \color{black}\vspace*{5pt}}
\setcounter{secnumdepth}{5}

\makeatletter
\renewcommand\@biblabel[1]{#1}
\renewcommand\@makefntext[1]%
{\noindent\makebox[0pt][r]{\@thefnmark\,}#1}
\makeatother
\renewcommand{\figurename}{Fig.}
\titleformat{\section}{\bfseries\normalsize}{\thesection.}{0.5em}{}
\titleformat{\subsection}{\bfseries\normalsize}{\thesubsection.}{0.5em}{}
\titleformat{\subsubsection}{\bfseries}{\thesubsubsection}{1em}{}
\setstretch{1.125}
\setlength{\skip\footins}{0.8cm}
\setlength{\footnotesep}{0.25cm}
\setlength{\jot}{10pt}
\titlespacing*{\section}{0pt}{3pt}{2pt}
\titlespacing*{\subsection}{0pt}{5pt}{1pt}
\setlength{\parskip}{0pt}
\setlength{\floatsep}{4pt plus 1pt minus 1pt}
\setlength{\textfloatsep}{4pt plus 1pt minus 1pt}
\setlength{\intextsep}{4pt plus 1pt minus 1pt}
\setlength{\dblfloatsep}{4pt plus 1pt minus 1pt}
\setlength{\dbltextfloatsep}{4pt plus 1pt minus 1pt}
\setlength{\abovecaptionskip}{2pt}
\setlength{\belowcaptionskip}{1pt}
\setlength{\abovedisplayskip}{4pt plus 1pt minus 1pt}
\setlength{\belowdisplayskip}{4pt plus 1pt minus 1pt}
\setlength{\abovedisplayshortskip}{2pt plus 1pt minus 1pt}
\setlength{\belowdisplayshortskip}{3pt plus 1pt minus 1pt}
\captionsetup{skip=2pt}
%%%END OF PAGE SETUP%%%

%%%FOOTER%%%
\fancyfoot{}
\fancyfoot[RO]{\footnotesize{\sffamily{1--\pageref{LastPage} ~\textbar  \hspace{2pt}\thepage}}}
\fancyfoot[LE]{\footnotesize{\sffamily{\thepage~\textbar\hspace{4.65cm} 1--\pageref{LastPage}}}}
\fancyhead{}
\renewcommand{\headrulewidth}{0pt}
\renewcommand{\footrulewidth}{0pt}
\setlength{\arrayrulewidth}{1pt}
\setlength{\columnsep}{6.5mm}
\setlength\bibsep{1pt}
%%%END OF FOOTER%%%

%%%FIGURE SETUP%%%
\makeatletter
\newlength{\figrulesep}
\setlength{\figrulesep}{0.5\textfloatsep}

\newcommand{\topfigrule}{\vspace*{-1pt}%
\noindent{\color{cream}\rule[-\figrulesep]{\columnwidth}{1.5pt}} }

\newcommand{\botfigrule}{\vspace*{-2pt}%
\noindent{\color{cream}\rule[\figrulesep]{\columnwidth}{1.5pt}} }

\newcommand{\dblfigrule}{\vspace*{-1pt}%
\noindent{\color{cream}\rule[-\figrulesep]{\textwidth}{1.5pt}} }

\makeatother
%%%END OF FIGURE SETUP%%%

%%%TITLE, AUTHORS AND ABSTRACT%%%
\twocolumn[
  \begin{@twocolumnfalse}
\vspace{0.4em}
\begin{tabular}{@{}m{4.5cm} p{13.5cm}@{}}

& \noindent\LARGE{\textbf{\shortstack[l]{Anisotropic Phonon Heat Flow and\\Thermoelectric Response in Tetragonal\\GeS$_{2}$ and GeSe$_{2}$}}} \\
\vspace{0.1cm} & \vspace{0.1cm} \\

& \noindent\Large{Neeraj Kulhari$^{1}$, Krishna Swaroop Sharma$^{1}$, K. C. Bhamu$^{2,*}$} \\
\vspace{0.05cm} & \vspace{0.05cm} \\

& \noindent\small{$^{1}$Department of Physics, IIS (Deemed to be University), Jaipur, Rajasthan 302020, India} \\
& \noindent\small{\shortstack[l]{$^{2}$Department of Physics, SLAS, Mody University of Science and Technology, Lakshmangarh,\\Sikar, Rajasthan 332311, India}} \\
& \noindent\small{$^{*}$Corresponding author: \href{mailto:kcbhamu85@gmail.com}{kcbhamu85@gmail.com}} \\
\vspace{0.1cm} & \vspace{0.1cm} \\

& \noindent\textbf{Abstract} \\
& \noindent\normalsize{The electronic structure, lattice dynamics, bonding, elastic response, and anisotropic thermoelectric transport properties of both tetragonal GeS$_2$ and GeSe$_2$ have been investigated using density functional theory, density functional perturbation theory, Wannier interpolation, and scattering-aware Boltzmann transport. The relaxed structures are mechanically and dynamically stable within the calculated harmonic description. The HSE03/Wannier band gaps are 2.48~eV for GeS$_2$ and 1.23~eV for GeSe$_2$, while substitution of S by Se lowers the upper phonon frequency from approximately 13.6 to 10.3~THz.

The phonon Boltzmann transport calculation reveals pronounced lattice-transport anisotropy. Within the relaxation-time approximation, the 300~K in-plane and cross-plane lattice thermal conductivities are 26.86 and 1.19~W~m$^{-1}$~K$^{-1}$ for GeS$_2$, and 18.74 and 1.52~W~m$^{-1}$~K$^{-1}$ for GeSe$_2$, respectively. These values decrease to 10.22 and 0.46~W~m$^{-1}$~K$^{-1}$ for GeS$_2$ and 7.25 and 0.58~W~m$^{-1}$~K$^{-1}$ for GeSe$_2$ at 800~K. Frequency-resolved analysis shows that most heat conduction is carried by low-frequency phonons, whereas the small cross-plane values reflect reduced out-of-plane phonon transport.

Combining the ShengBTE RTA lattice tensors with AMSET electronic coefficients gives a $zT$ value of 0.257 for n-type cross-plane GeS$_2$ at 800~K and $10^{19}$~cm$^{-3}$. The corresponding PBE-AMSET estimate for GeSe$_2$ gives a value of 0.066 for p-type cross-plane transport at 800~K and $3\times10^{20}$~cm$^{-3}$. LOBSTER analysis identifies mixed covalent--ionic Ge--X bonding, with Ge--S bonds having a larger stabilizing ICOHP magnitude than Ge--Se bonds ($-5.27$ versus $-4.74$~eV per bond). These results show that tetragonal GeX$_2$ compounds are strongly anisotropic thermoelectrics with moderate calculated $zT$ values, and that their cross-plane response benefits from suppressed lattice heat transport.} \\
& \noindent\textbf{Keywords:} first-principles calculations; germanium chalcogenides; lattice thermal conductivity; ShengBTE; AMSET; thermoelectric transport \\

\end{tabular}
\end{@twocolumnfalse}\vspace{0.25cm}
]

%%%END OF TITLE, AUTHORS AND ABSTRACT%%%

%%%FONT SETUP%%%
\renewcommand*\rmdefault{ptm}\normalfont\upshape
\rmfamily
\vspace{-0.25cm}

%%%MAIN TEXT%%%%
\section{Introduction}

Growing energy demand and continued dependence on non-renewable resources have increased the need for clean energy-conversion materials.\cite{Aswal2016TEG,Wei2020Review,HeTritt2017Science} Thermoelectric materials are of interest because they directly convert heat into electricity and can also provide solid-state cooling, making them useful for waste-heat recovery in industrial, automotive, and other dissipative systems.\cite{Aswal2016TEG,Snyder2008ComplexTE}

The performance of a thermoelectric material is commonly quantified by the dimensionless figure of merit, $zT$, defined as
\begin{equation}
 zT = \frac{\sigma S^{2} T}{\kappa_e + \kappa_l},
 \label{eq:zt}
\end{equation}
where $\sigma$ is the electrical conductivity, $S$ is the Seebeck coefficient, $T$ is the absolute temperature, and $\kappa_e$ and $\kappa_l$ are the electronic and lattice contributions to thermal conductivity, respectively.\cite{Snyder2008ComplexTE,Duong2016SnSe,ZhaoNature2014SnSe} High $zT$ requires a large power factor, $\sigma S^{2}$, and low total thermal conductivity. These quantities are mutually coupled: increasing carrier concentration can improve $\sigma$ but often raises $\kappa_e$ through the Wiedemann--Franz relation, while lower carrier scattering may also increase heat transport.\cite{Wei2020Review,Zhu2022PbSe,Chen2017PbTeDislocations,Wu2019LatticeStrain} By contrast, $\kappa_l$ can often be reduced more independently by increasing phonon scattering through alloying, nanostructuring, dislocations, point defects, or hierarchical microstructures.\cite{Zhu2022PbSe,Perumal2017GeTe,ZhangPNAS2013SnTe,Wu2019LatticeStrain}

These design principles have been implemented across diverse thermoelectric material classes, including skutterudites, half-Heusler compounds, chalcogenides, and layered low-dimensional systems.\cite{Wei2020Review,AryanaMVB2024Chalcogenides} Among them, IV--VI chalcogenides such as SnSe and SnTe have demonstrated record-high $zT$ values through coordinated band and phonon engineering.\cite{Duong2016SnSe,ZhaoNature2014SnSe,Perumal2017GeTe,ZhangPNAS2013SnTe,Zhu2022PbSe} Despite these successes, several widely used materials still face practical challenges: Bi$_2$Te$_3$-based alloys, for example, exhibit performance degradation at elevated temperatures, which lowers useful $zT$, whereas skutterudites and many half-Heusler compounds possess intrinsically high lattice thermal conductivity.\cite{Wei2020Review,Aswal2016TEG} These limitations motivate continued exploration of non-toxic, earth-abundant compounds that combine favorable electronic properties with intrinsically low lattice thermal conductivity.\cite{AryanaMVB2024Chalcogenides}

Layered and low-dimensional materials are promising thermoelectric candidates because structural anisotropy can influence, and partly decouple, electronic and phonon transport.\cite{Li2021Anisotropic2DReview,Zhao2020Anisotropic2D,Du2021SymmetryBreaking} Group IV--VI chalcogenides containing Ge or Sn, such as GeS, GeSe, SnS, and SnSe, combine useful band structures with relatively abundant and environmentally compatible elements.\cite{Chen2022GeSMicroribbons,Li2021Anisotropic2DReview} Germanium dichalcogenides, GeX$_2$ (X = S, Se), have historically been studied mainly as glasses and for phase-change or resistive-switching applications. Recent work on metavalent bonding in main-group chalcogenides helps explain how these compounds can combine favorable electronic transport with strongly anharmonic phonon behavior, while their crystalline layered forms are now attracting interest for optoelectronic and nanoelectronic devices.\cite{AryanaMVB2024Chalcogenides} Experiments have shown strong in-plane structural, vibrational, and optical anisotropy in exfoliated GeS$_2$ and GeSe$_2$ nanoflakes, and ultrathin $\beta$-GeSe$_2$ flakes grown by van der Waals epitaxy show promising polarization-sensitive photodetection.\cite{Wang2020GeS2Anisotropy,Shafirin2026GeS2Photonics,Chen2020GaSeMemory,Yang2019WeakInterlayerGeSe2}

Recent studies also highlight the optical and transport potential of GeS$_2$. Atomic-scale characterization of monoclinic layered GeS$_2$ nanostructures has linked strong in-plane anisotropy to anisotropic electronic and optical properties.\cite{Wang2020GeS2Anisotropy} Layered GeS$_2$ has also been reported to exhibit an exceptionally high refractive index and strong optical anisotropy in the blue and near-ultraviolet frequency ranges, making it relevant for short-wavelength nanophotonics and metasurfaces.\cite{Shafirin2026GeS2Photonics} First-principles studies have further examined adsorption, electronic, and optical properties of GeS$_2$ monolayers, as well as the dependence of calculated 2D band gaps on the chosen exchange--correlation functionals.\cite{Gao2022GeS2Gas,Tran2021Bandgap2D,Tse2024GeS2HSE03}

Beyond the monoclinic layered phases, GeS$_2$ also occurs in a tetragonal framework (t-GeS$_2$) with space group $P4_{2}/nmc$, where corner-sharing GeS$_4$ tetrahedra form quasi-two-dimensional sheets oriented along the $(0,0,1)$ direction.\cite{MPGeS2Tetragonal,GeS2C44Structure} This polymorph is a useful platform for connecting wide-gap semiconducting behavior, tetrahedral coordination, and anisotropic bonding with optical and transport properties. However, the electronic, optical, and phonon-transport properties of bulk t-GeS$_2$ remain comparatively underexplored from a thermoelectric perspective.\cite{Wang2020GeS2Anisotropy,Shafirin2026GeS2Photonics} The related GeSe$_2$ system is also less understood, especially in terms of anisotropic thermoelectric response. Prior work on anisotropic layered GeSe$_2$ reported a very small cleavage energy of about 0.05~J~m$^{-2}$ and weak layer-dependent changes in the electronic and Raman spectra, showing that weak interlayer coupling is a realistic structural feature of this chemistry.\cite{Yang2019WeakInterlayerGeSe2}

In this work, the electronic structure and lattice thermal transport of bulk t-GeS$_2$ are investigated and compared with t-GeSe$_2$ as a structural and chemical analogue. For t-GeS$_2$, we use the tetragonal reference structure with $a = b = 3.50$~\AA{}, $c = 11.20$~\AA{}, $\alpha = \beta = \gamma = 90.00^\circ$, Ge atoms at $(0.25,\,0.75,\,0.25)$ and $(0.75,\,0.25,\,0.75)$, and S atoms at $(0.25,\,0.25,\,0.37686)$, $(0.75,\,0.75,\,0.62314)$, $(0.25,\,0.25,\,0.87686)$, and $(0.75,\,0.75,\,0.12314)$.\cite{Wang2020GeS2Anisotropy} For t-GeSe$_2$, the substituted and relaxed structure used in the calculations has $a=b=3.703$~\AA{} and $c=11.284$~\AA{}, with $\alpha=\beta=\gamma=90^\circ$; the Ge atoms occupy the same fractional positions, while Se atoms are located at $(0.25,\,0.25,\,0.38604)$, $(0.75,\,0.75,\,0.61396)$, $(0.25,\,0.25,\,0.88604)$, and $(0.75,\,0.75,\,0.11396)$. DFT and DFPT are used to obtain the electronic band structure, density of states, and elastic properties. Electronic transport is evaluated with state-dependent acoustic-deformation-potential, polar-optical-phonon, and ionized-impurity scattering,\cite{Ganose2021AMSET,Claes2025PhononTransport,Zhou2026TransportReview} while lattice thermal conductivity is obtained with ShengBTE.\cite{Li2014ShengBTE} Band-structure analysis, ICOOP, and ICOHP descriptors are then used to clarify the microscopic origin of the electronic and phonon transport behavior.

\section{Computational details}
\subsection{Electronic structure and transport calculations}

The electronic-structure calculations were carried out within Kohn--Sham density functional theory (DFT) using the plane-wave pseudopotential framework implemented in \textsc{Quantum ESPRESSO}.\cite{Giannozzi2009QE,Giannozzi2020QE} We focused on the tetragonal GeS$_2$ phase (t-GeS$_2$) with space group $P4_{2}/nmc$. The reported reference structure has $a=b=3.50$~\AA{} and $c=11.20$~\AA{}, with $\alpha=\beta=\gamma=90^\circ$.\cite{Wang2020GeS2Anisotropy} In the production \textsc{Quantum ESPRESSO} inputs used here, the tetragonal cell was defined by \texttt{ibrav = 6}, \texttt{celldm(1) = 6.636200}, and \texttt{celldm(3) = 3.133016}, corresponding to $a=b=3.512$~\AA{} and $c=11.002$~\AA{}, with two Ge and four S atoms per primitive cell.

Internal-coordinate optimization at the fixed tetragonal cell and ground-state electronic properties were obtained using the generalized gradient approximation (GGA) in the Perdew--Burke--Ernzerhof (PBE) formulation for the exchange--correlation functional,\cite{Perdew1996PBE} combined with a semi-empirical Grimme D2 dispersion correction (\texttt{vdw\_corr = `grimme-d2'}) to account for van der Waals interactions.\cite{Grimme2006D2,Grimme2010D3} Except for the separate \textsc{LOBSTER} projection step discussed below, all \textsc{Quantum ESPRESSO} calculations used scalar-relativistic ultrasoft RRKJ pseudopotentials from the PSLibrary (\nolinkurl{Ge.pbe-n-rrkjus_psl.1.0.0.UPF}, \nolinkurl{S.pbe-n-rrkjus_psl.1.0.0.UPF}, and \nolinkurl{Se.pbe-n-rrkjus_psl.1.0.0.UPF}).\cite{DalCorso2014PSLibrary} This includes the PBE structural, elastic, band-structure, DFPT, deformation-potential, and AMSET-input calculations, as well as the HSE03 band-reference calculations. Keeping the same ultrasoft pseudopotential family across the main workflow avoids mixing structural, electronic, and response properties from different pseudopotential descriptions. The plane-wave kinetic-energy cutoff and charge-density cutoff were set to \texttt{ecutwfc = 60~Ry} and \texttt{ecutrho = 480~Ry}, respectively, and Brillouin-zone integrations were performed using an $8\times 8\times 4$ Monkhorst--Pack $k$-point mesh with \texttt{occupations = `fixed'}. Self-consistent PBE calculations employed a convergence threshold of \texttt{conv\_thr} = $1.0\times 10^{-10}$~Ry on the total energy, with \texttt{mixing\_beta = 0.15} and a maximum of 200 electronic iterations.

To obtain more reliable band gaps for interpreting carrier excitation and transport trends, we performed additional self-consistent calculations using the screened HSE03 hybrid functional, as implemented in \textsc{Quantum ESPRESSO}.\cite{Heyd2003HSE03,Paier2006HSEsolids,Krukau2006HSEScreening} In these calculations, the short-range Hartree--Fock exchange fraction and screening parameter were set to \texttt{exx\_fraction = 0.25} and \texttt{screening\_parameter = 0.106}, respectively, with the $\mathbf{q}$-point mesh for exchange integrals defined by \texttt{nqx1 = 2}, \texttt{nqx2 = 2}, and \texttt{nqx3 = 1}. The same relaxed internal coordinates were used, and the Brillouin zone was sampled with a slightly coarser $6\times 6\times 2$ $k$-point grid (\texttt{K\_POINTS \{automatic\} 6 6 2 0 0 0}). The electronic convergence threshold for HSE03 was set to \texttt{conv\_thr} = $1.0\times 10^{-8}$~Ry with \texttt{mixing\_beta = 0.10} and up to 300 electronic iterations. This staged PBE-then-HSE03 workflow, in which cheaper GGA relaxations are followed by a single-point hybrid-functional correction, is a common strategy for obtaining improved semiconductor band gaps without the full cost of hybrid-functional geometry optimization, and has been used previously for related GeS$_2$ systems.\cite{Tse2024GeS2HSE03,Tran2021Bandgap2D}

For visualization, the PBE and HSE03 bands were interpolated with \textsc{Wannier90}.\cite{Marzari1997MLWF,Souza2001Wannier,Yates2007WannierInterpolation,Mostofi2008Wannier90,Pizzi2020Wannier90} Electronic transport was calculated with \textsc{AMSET} v0.5.1, using its state-dependent Boltzmann transport formalism.\cite{Ganose2021AMSET} Because AMSET does not natively read Quantum ESPRESSO outputs, a Python interface based on the \texttt{PWxml} parser in \textsc{pymatgen} was used to extract the structure, $k$ points, eigenvalues, and electron count from \texttt{data-file-schema.xml} and write the AMSET \texttt{band\_structure\_data.json} input.\cite{Ong2013Pymatgen} Orbital projections were represented by placeholder weights, so the wavefunction-overlap factors are approximate. ADP, POP, and IMP scattering were included using the calculated elastic, dielectric, deformation-potential, and optical-phonon inputs. Calculations covered 300--800~K and $10^{18}$--$10^{22}$~cm$^{-3}$ on dense AMSET interpolation meshes of $81\times81\times25$ for GeS$_2$ and $81\times81\times27$ for GeSe$_2$. These meshes are post-processing grids generated from the converted dense NSCF band structures and are not the same as the Quantum ESPRESSO SCF/NSCF $k$ meshes. The converted PBE transport bands yield a 0.99~eV GeS$_2$ gap, while GeSe$_2$ lies very close to gap closure on the dense transport mesh, with a small $\sim0.015$~eV band overlap; this near-zero-gap PBE limit restricts quantitative high-temperature interpretation for GeSe$_2$.\cite{Claes2025PhononTransport,Zhou2026TransportReview}
As an independent constant-relaxation-time reference, the HSE03/Wannier bands were also post-processed with \textsc{BoltzTraP2}; these chemical-potential-dependent trends are included in the Supplementary Information and are not used for the quantitative $zT$ values reported below.\cite{Madsen2018BoltzTraP2}

Harmonic lattice dynamics were investigated with DFPT as implemented in the \textsc{PHonon} component of \textsc{Quantum ESPRESSO}.\cite{Giannozzi2009QE,Giannozzi2020QE,Baroni2001DFPT} The PBE+DFT-D2 ground state used the same ultrasoft pseudopotentials, an $8\times8\times4$ electronic $k$ mesh, and wave-function and charge-density cutoffs of 60 and 480~Ry, respectively. This choice is also efficient for response calculations involving dense $q$-point sampling, Born effective charges, and dielectric tensors while retaining the same PBE+D2 structural description. Dynamical matrices were evaluated on a $20\times20\times6$ $q$ mesh and transformed to real-space harmonic force constants with \texttt{q2r.x}. Born effective charges and the high-frequency dielectric tensor were obtained at $\Gamma$ using \texttt{epsil=.true.}; the corresponding non-analytical dipole--dipole correction was retained when interpolating the dispersion with \texttt{matdyn.x}.\cite{GonzeLee1997DFPT} The crystal acoustic sum rule was imposed, and the total phonon DOS was evaluated by tetrahedron integration on a $40\times40\times20$ uniform mesh with a 1~cm$^{-1}$ energy spacing. An independent $4\times4\times2$ finite-displacement calculation with \textsc{Phono3py}, supplied with the same dielectric tensor and Born charges, was used to verify the harmonic spectrum and the non-analytic splitting near $\Gamma$.\cite{Togo2023Phonopy,Togo2023Implementation}

The lattice thermal conductivity was subsequently calculated from second- and third-order force constants using the phonon Boltzmann transport equation implemented in \textsc{ShengBTE}.\cite{Li2014ShengBTE} Third-order force constants were generated using $4\times4\times2$ finite-displacement supercells with \texttt{thirdorder\_espresso.py}; after the force calculations, \texttt{thirdorder\_espresso.py scf.in reap 4 4 2 -7} was used for both compounds to retain anharmonic interactions through the seventh-nearest-neighbor shell. Kinetic-energy cutoffs of 65 and 520~Ry were used for the wave functions and charge density, and the Grimme D2 dispersion correction was included consistently to describe the interlayer interaction.\cite{Grimme2006D2,Grimme2010D3} The transport equation was sampled on a $20\times20\times6$ phonon wave-vector mesh; both RTA and iterative tensors were checked, and the RTA tensor was used for the internally consistent $zT$ estimates reported below.

\section{Results and discussion}
\subsection{Crystal structures}

\begin{figure}[!htbp]
\centering

\begin{subfigure}[t]{0.46\columnwidth}
    \centering
    \includegraphics[width=\linewidth,height=0.115\textheight,keepaspectratio]{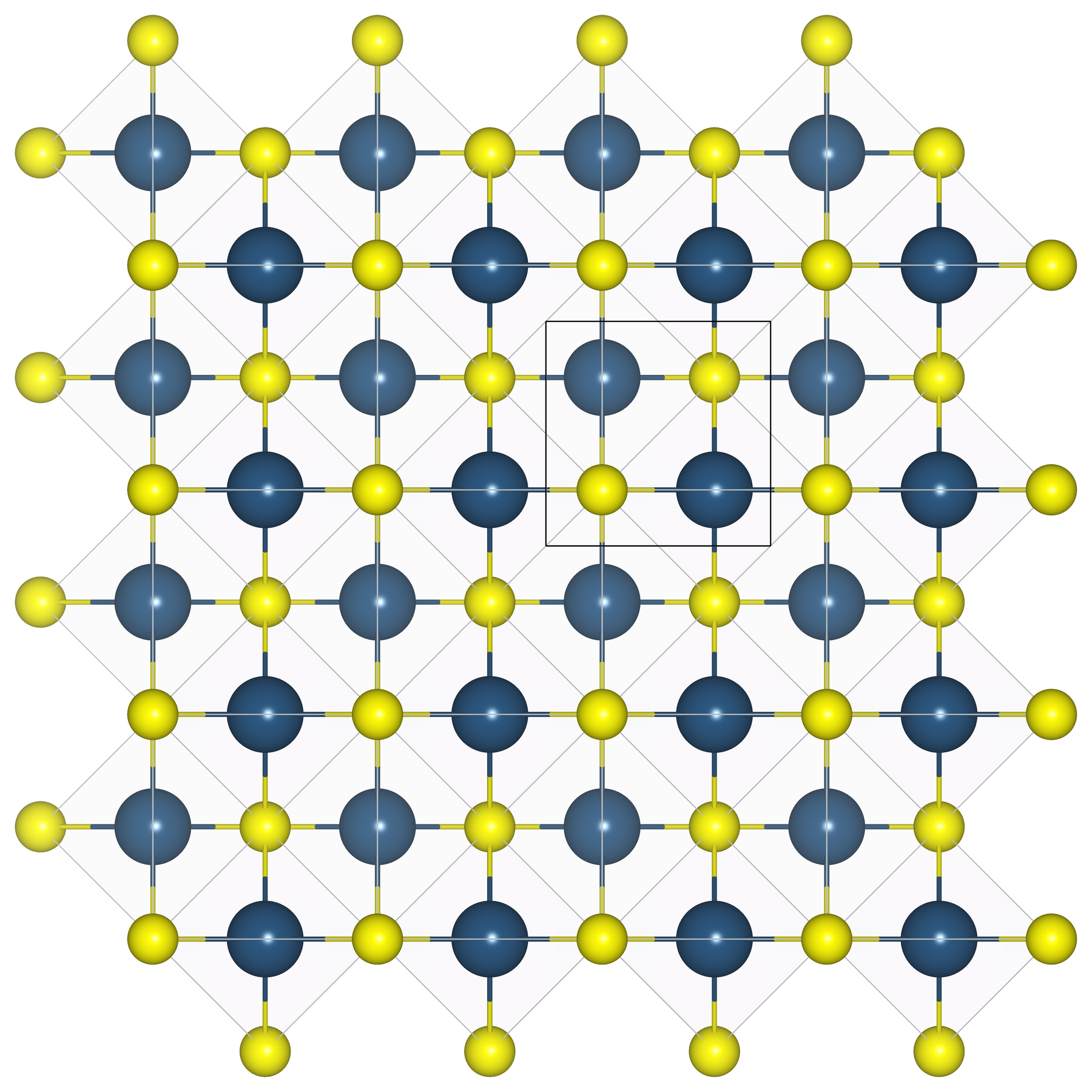}
    \caption{Top view}
\end{subfigure}
\hfill
\begin{subfigure}[t]{0.40\columnwidth}
    \centering
    \includegraphics[width=\linewidth,height=0.115\textheight,keepaspectratio]{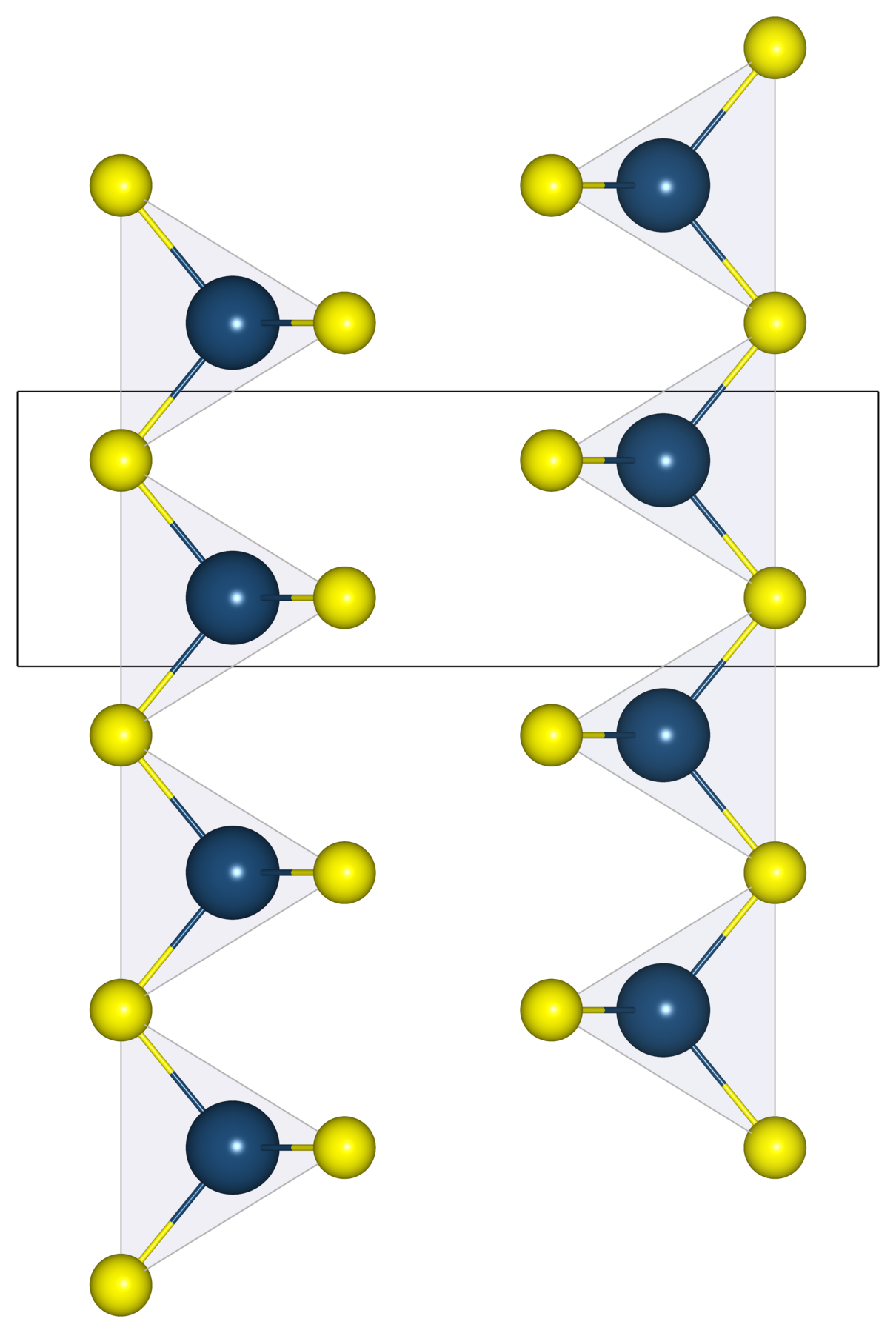}
    \caption{Side view}
\end{subfigure}

\caption{
Crystal structure of tetragonal GeS$_2$ (space group P4$_2$/nmc): (a) top view and (b) side view. Blue and yellow spheres denote Ge and S atoms, respectively; translucent polyhedra in panel (a) show GeS$_4$ tetrahedra, and the projected unit cell is indicated in panel (b).
}
\label{fig:GeS2_structure}

\end{figure}

\begin{figure}[!htbp]
\centering

\begin{subfigure}[t]{0.46\columnwidth}
    \centering
    \includegraphics[width=\linewidth,height=0.115\textheight,keepaspectratio]{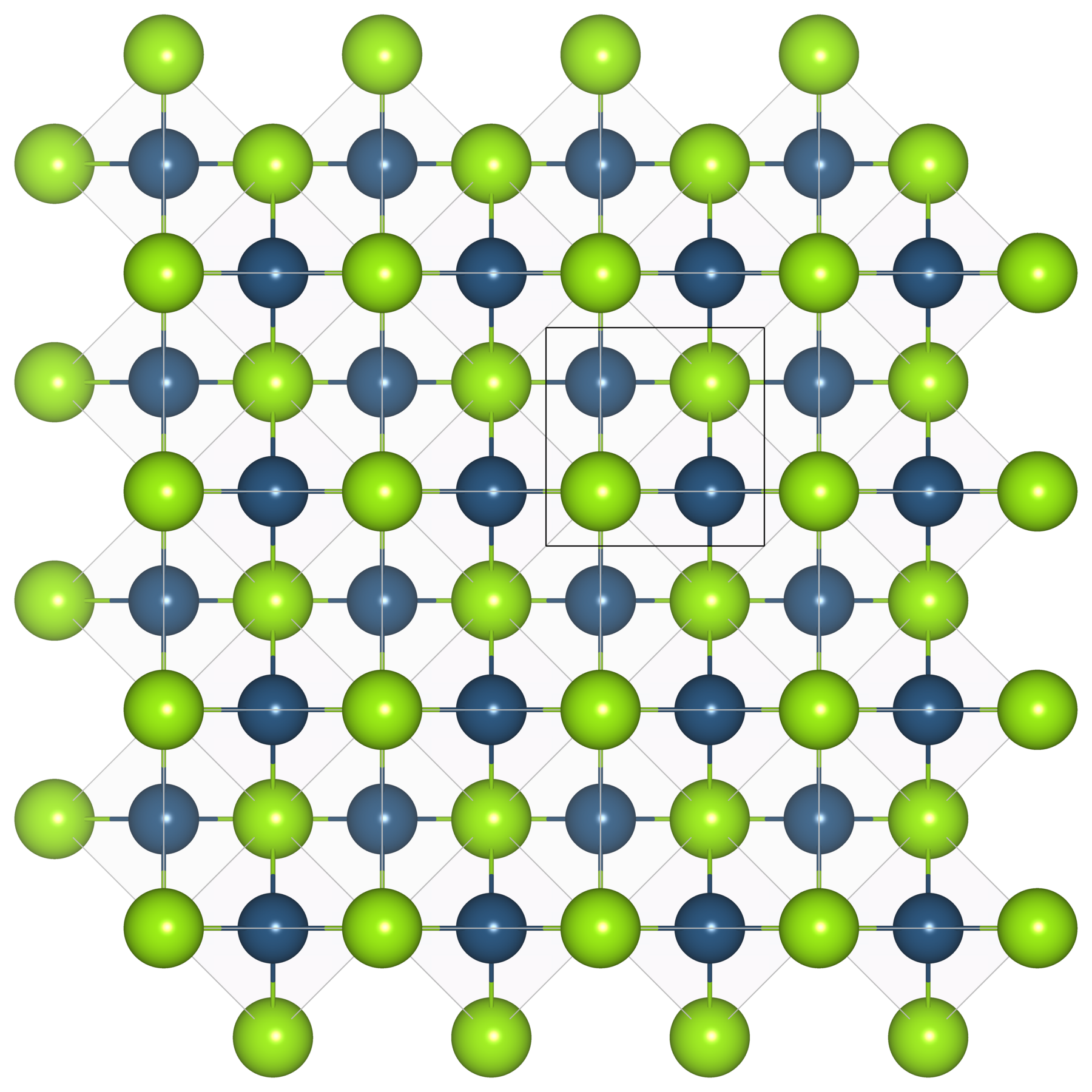}
    \caption{Top view}
\end{subfigure}
\hfill
\begin{subfigure}[t]{0.40\columnwidth}
    \centering
    \includegraphics[width=\linewidth,height=0.115\textheight,keepaspectratio]{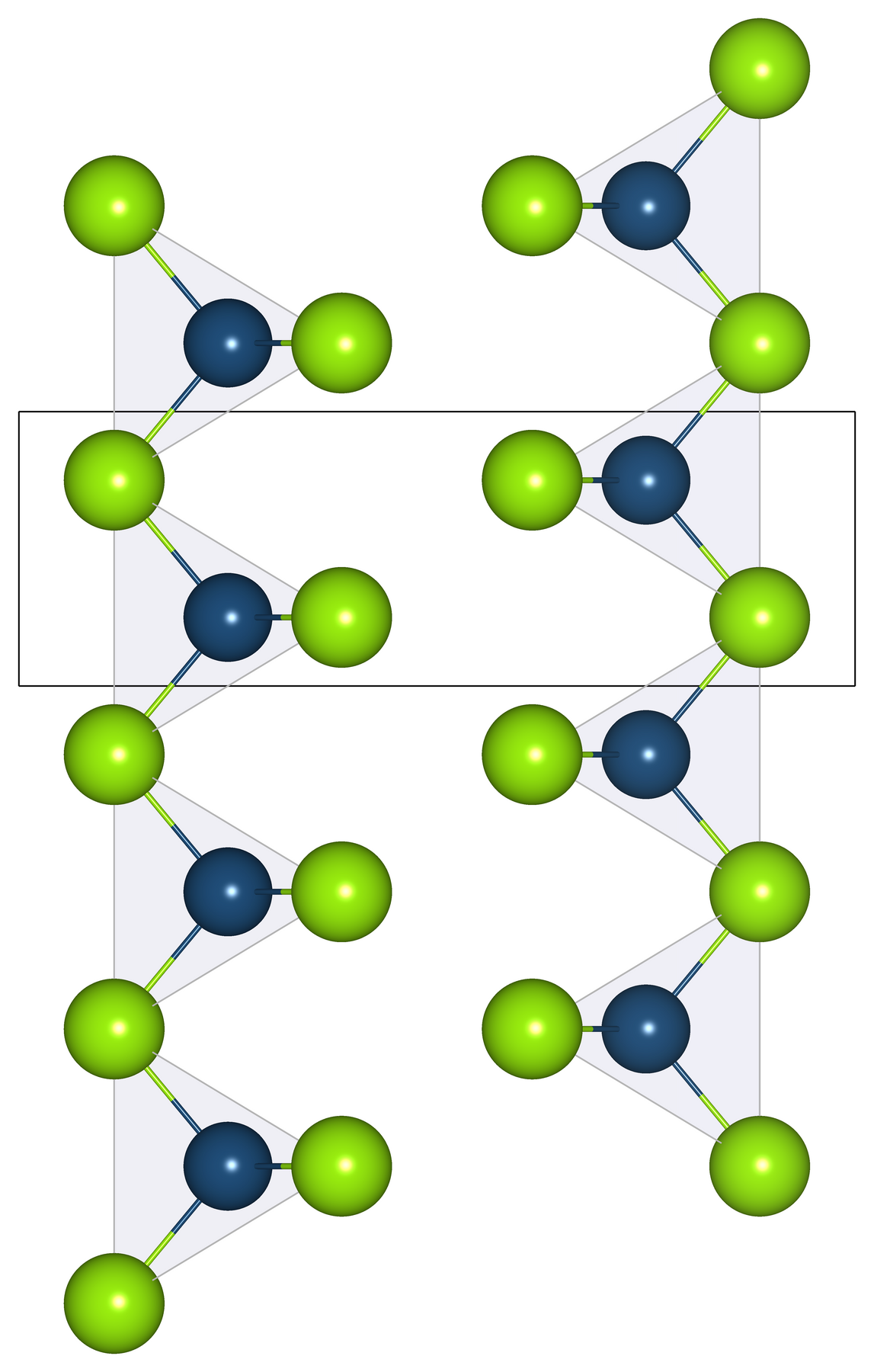}
    \caption{Side view}
\end{subfigure}
\caption{
Crystal structure of tetragonal GeSe$_2$ (space group P4$_2$/nmc): (a) top view and (b) side view. Blue and green spheres denote Ge and Se atoms, respectively; translucent polyhedra in panel (a) show GeSe$_4$ tetrahedra, and the projected unit cell is indicated in panel (b).
}
\label{fig:GeSe2_structure}

\end{figure}

GeX$_2$ (X = S, Se) crystallizes in a tetragonal structure with space group $P4_{2}/nmc$.\cite{MPGeS2Tetragonal,GeS2C44Structure} The calculations were based on the reported tetragonal GeS$_2$ reference lattice, $a=b=3.50$~\AA{} and $c=11.20$~\AA{}, together with the corresponding internal coordinates.\cite{Wang2020GeS2Anisotropy} The cell used in our GeS$_2$ SCF input is $a=b=3.512$~\AA{} and $c=11.002$~\AA{}, obtained from \texttt{celldm(1) = 6.636200} and \texttt{celldm(3) = 3.133016}. These values remain close to the reported tetragonal reference structure.\cite{MPGeS2Tetragonal,GeS2C44Structure}

The tetragonal GeSe$_2$ structure was generated by replacing S with Se in the same GeX$_2$ framework and using the corresponding SCF cell defined by \texttt{celldm(1) = 6.998187741} and \texttt{celldm(3) = 3.047055436}. This gives $a=b=3.703$~\AA{} and $c=11.284$~\AA{}, with $\alpha=\beta=\gamma=90^\circ$. The larger lattice constants are consistent with the larger ionic size of Se compared with S. Because detailed experimental lattice parameters for GeSe$_2$ in the $P4_{2}/nmc$ structure have not yet been reported, this geometry can serve as a first-principles reference for future studies. The final variable-cell relaxations gave residual pressures of only $-0.03$~kbar for GeS$_2$ and $-0.01$~kbar for GeSe$_2$; subsequent fixed-cell SCF calculations also remained close to zero pressure, with 0.13 and $-0.02$~kbar, respectively.

Figures~\ref{fig:GeS2_structure} and \ref{fig:GeSe2_structure} show that both compounds consist of corner-sharing GeX$_4$ tetrahedra arranged into layered GeX$_2$ sheets along the $(0,0,1)$ direction. Each Ge$^{4+}$ cation is coordinated by four chalcogen anions, forming a quasi-two-dimensional tetrahedral network analogous to a single slab of the high-pressure, layer-structured, tetragonal HgI$_2$-type phase.\cite{GeS2C44Structure} The top views emphasize the in-plane periodic arrangement of Ge and chalcogen atoms, while the side views show the stacking of the GeX$_2$ layers along the $c$ axis.

\subsection{Elastic properties}

Elastic properties provide a direct measure of mechanical stability, stiffness, bonding anisotropy, and deformation response. The elastic constants were calculated with \textsc{thermo\_pw}, using the standard finite-strain approach interfaced with \textsc{Quantum ESPRESSO}.\cite{Giannozzi2009QE,Giannozzi2020QE,Malica2021ThermoPWElastic} For both compounds, the final static elastic calculations used \texttt{frozen\_ions = .false.}, allowing the internal atomic coordinates to relax under each imposed strain. GeS$_2$ was fitted using six strain geometries, a strain amplitude of 0.005, and a third-order polynomial. For the softer GeSe$_2$ lattice, a smaller strain amplitude of 0.002 with four strain geometries and a second-order polynomial was used to avoid unstable internal-coordinate relaxations. For tetragonal crystals, the stiffness tensor has six independent second-order elastic constants: $C_{11}$, $C_{12}$, $C_{13}$, $C_{33}$, $C_{44}$, and $C_{66}$. The calculated values are listed in Table~\ref{tab:elastic_constants}. All values are reported in GPa after converting the \textsc{thermo\_pw} output from kbar using $1~\mathrm{kbar}=0.1~\mathrm{GPa}$.

\begin{table}[H]
\centering
\caption{Calculated relaxed-ion elastic constants of tetragonal GeS$_2$ and GeSe$_2$ in GPa.}
\label{tab:elastic_constants}
\footnotesize
\setlength{\tabcolsep}{3pt}
\begin{tabular}{lcc}
\hline
Elastic constant & GeS$_2$ & GeSe$_2$ \\
\hline
$C_{11}$ & 150.67 & 124.05 \\
$C_{12}$ & -2.26 & 4.60 \\
$C_{13}$ & 13.93 & 20.92 \\
$C_{33}$ & 27.48 & 37.32 \\
$C_{44}$ & 19.70 & 27.04 \\
$C_{66}$ & 3.49 & 2.68 \\
\hline
\end{tabular}
\end{table}

The elastic constants satisfy the Born mechanical stability criteria for a tetragonal crystal: $C_{11}>|C_{12}|$, $2C_{13}^{2}<C_{33}(C_{11}+C_{12})$, $C_{44}>0$, and $C_{66}>0$.\cite{Mouhat2014ElasticStability} This confirms that the optimized tetragonal GeS$_2$ and GeSe$_2$ structures are mechanically stable against small homogeneous strains. Both compounds show strongly anisotropic stiffness: the in-plane constants are large, with $C_{11}=150.67$~GPa for GeS$_2$ and 124.05~GPa for GeSe$_2$, whereas the out-of-plane constants are much smaller, $C_{33}=27.48$ and 37.32~GPa, respectively. This contrast is consistent with the layered tetragonal framework and weaker resistance to deformation along the stacking direction.

The small $C_{66}$ values indicate that both compounds are especially soft against in-plane shear deformation, even though their longitudinal and out-of-plane compression constants are much larger. This shear softness is important because soft deformation channels are often associated with anisotropic bonding, low phonon group velocities, and enhanced phonon scattering, all of which are favorable for reducing lattice thermal conductivity. The $C_{12}$ values are small compared with $C_{11}$, indicating weak coupling between orthogonal in-plane normal strains.

\begin{table}[H]
\centering
\caption{Voigt, Reuss, and Hill averaged relaxed-ion mechanical parameters of tetragonal GeS$_2$ and GeSe$_2$.}
\label{tab:elastic_moduli}
\footnotesize
\setlength{\tabcolsep}{3pt}
\begin{tabular}{lcc}
\hline
Parameter & GeS$_2$ & GeSe$_2$ \\
\hline
$B_V$ (GPa) & 42.23 & 42.04 \\
$B_R$ (GPa) & 25.00 & 32.82 \\
$B_H$ (GPa) & 33.61 & 37.43 \\
$G_V$ (GPa) & 28.79 & 27.28 \\
$G_R$ (GPa) & 10.64 & 9.46 \\
$G_H$ (GPa) & 19.72 & 18.37 \\
$E_V$ (GPa) & 70.38 & 67.29 \\
$E_R$ (GPa) & 27.96 & 25.90 \\
$E_H$ (GPa) & 49.17 & 46.60 \\
$\nu_V$ & 0.222 & 0.233 \\
$\nu_R$ & 0.314 & 0.368 \\
$\nu_H$ & 0.247 & 0.268 \\
\hline
\end{tabular}
\end{table}

The Hill averages, which are the arithmetic means of the Voigt and Reuss bounds, give $B=33.61$~GPa, $G=19.72$~GPa, and $E=49.17$~GPa for GeS$_2$.\cite{Hill1952Elastic} The corresponding values for GeSe$_2$ are $B=37.43$~GPa, $G=18.37$~GPa, and $E=46.60$~GPa. These moderate values show that both tetragonal compounds are mechanically soft compared with densely bonded three-dimensional covalent solids. The Pugh ratio, $B/G$, is 1.70 for GeS$_2$ and 2.04 for GeSe$_2$.\cite{Pugh1954Elastic} Thus GeS$_2$ lies slightly below the empirical ductile--brittle threshold of 1.75, while GeSe$_2$ lies above it, indicating a somewhat more ductile average response.

The sound velocities and Debye temperatures obtained from the Voigt--Reuss--Hill elastic averages provide an additional indicator of heat transport. In the Slack model, the lattice thermal conductivity scales strongly with the Debye temperature, approximately through a $\theta_D^3$ dependence.\cite{Slack1979ThermalConductivity} For GeS$_2$, \textsc{thermo\_pw} gives bulk and shear sound velocities of 3.169 and 2.427~km~s$^{-1}$, respectively, an average Debye sound velocity of 2.462~km~s$^{-1}$, and a Debye temperature of 259.2~K. For GeSe$_2$, the corresponding values are 2.750, 1.927, and 2.047~km~s$^{-1}$, with a Debye temperature of 206.3~K. These low sound velocities and Debye temperatures are consistent with soft acoustic modes and support the low lattice thermal conductivity obtained from ShengBTE.

Overall, the elastic response confirms that both tetragonal compounds are mechanically stable and anisotropic, with pronounced shear softness that supports low lattice thermal conductivity and anisotropic thermoelectric behavior.

\subsection{Phonon dispersion and vibrational density of states}

\begin{figure}[t]
\centering
\includegraphics[width=\columnwidth]{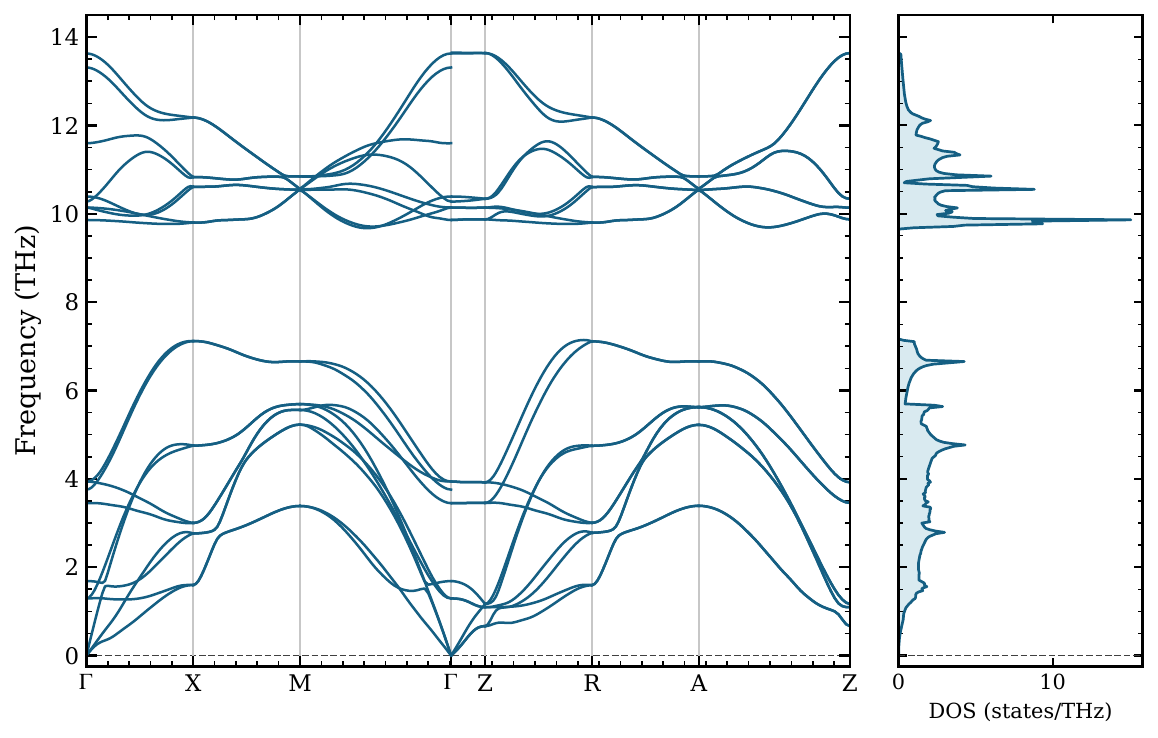}
\caption{DFPT phonon dispersion and total vibrational density of states of tetragonal GeS$_2$. Frequencies are interpolated along $\Gamma$--X--M--$\Gamma$--Z--R--A--Z, while the DOS is integrated on a $40\times40\times20$ uniform mesh. The non-analytical correction from the calculated dielectric tensor and Born effective charges is included.}
\label{fig:GeS2_phonon_dos}
\end{figure}

\begin{figure}[t]
\centering
\includegraphics[width=\columnwidth]{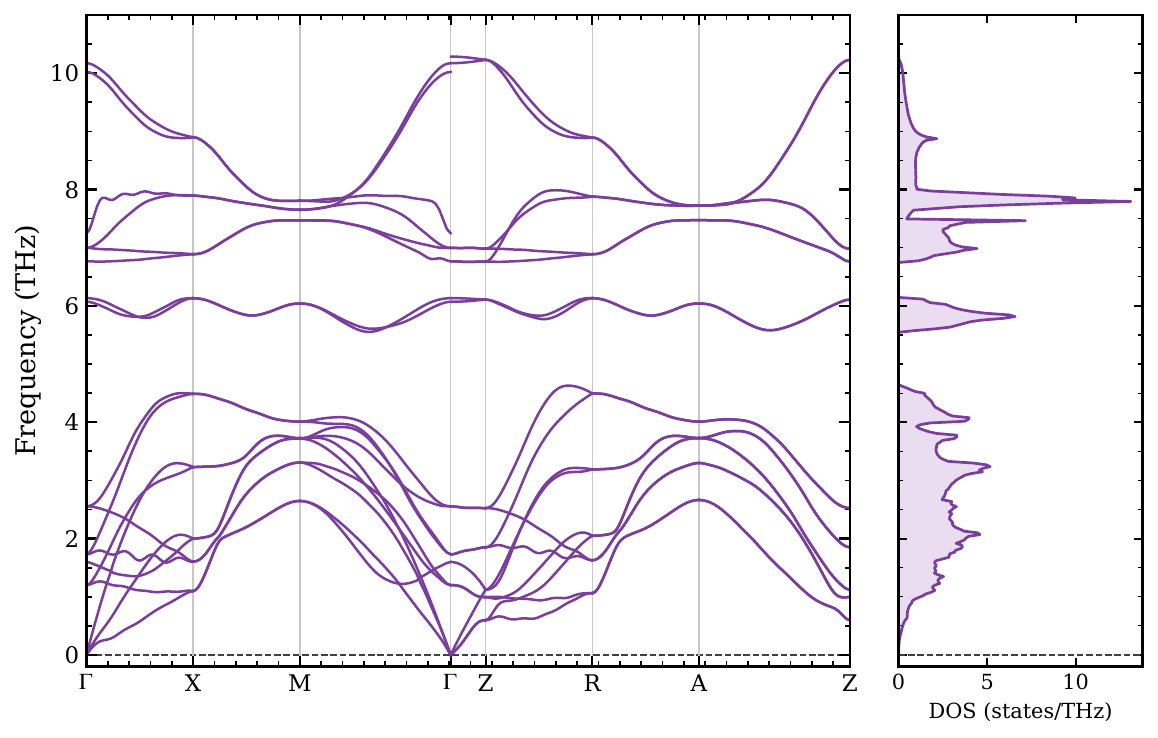}
\caption{DFPT phonon dispersion and total vibrational density of states of tetragonal GeSe$_2$, calculated with the same path, non-analytical correction, and $40\times40\times20$ DOS mesh used for GeS$_2$.}
\label{fig:GeSe2_phonon_dos}
\end{figure}

Figure~\ref{fig:GeS2_phonon_dos} contains 18 branches, as required for the six-atom primitive cell: three acoustic and 15 optical modes. No significant imaginary branch is observed along the sampled path; any tiny numerical deviation near $\Gamma$, if present, is within the interpolation tolerance. The relaxed tetragonal GeS$_2$ phase is therefore harmonically dynamically stable. This conclusion is complementary to the elastic stability conditions because the latter test only homogeneous strain, whereas the phonon calculation probes collective atomic displacements throughout the Brillouin zone. Similar use of phonon spectra to establish stability has been reported for other two-dimensional Ge--S phases, although their structures and dimensionality differ from the bulk tetragonal phase studied here.\cite{Ali2024GeS2Phonons}

The spectrum separates into a lower manifold extending to approximately 7.2~THz and a higher optical manifold between about 9.7 and 13.6~THz, leaving a gap of roughly 2.5~THz. The comparatively dispersive low-frequency branches provide the main harmonic heat-carrying channels, whereas several optical branches are relatively flat and therefore have small group velocities. Their weak dispersion produces pronounced DOS maxima, particularly near 9.8--10.7~THz. Recent polarization-resolved Raman measurements and first-principles mode analysis of layered GeS$_2$ assign the lower-frequency region mainly to Ge--S bending and tetrahedral breathing motions and the 340--450~cm$^{-1}$ interval (approximately 10.2--13.5~THz) primarily to S-dominated Ge--S stretching.\cite{Slavich2025GeS2Vibrations} This assignment is consistent with the isolated high-frequency manifold and its upper limit near 13.6~THz in Fig.~\ref{fig:GeS2_phonon_dos}.

Near $\Gamma$, selected optical frequencies depend on whether the wave vector approaches in plane or along $\Gamma$--Z. This is the expected non-analytic long-wavelength response produced by the anisotropic Born effective charges and dielectric screening, rather than an unstable or poorly converged mode.\cite{Baroni2001DFPT,GonzeLee1997DFPT} The two limiting approaches are therefore shown as separate path segments instead of being joined by an artificial vertical line. Finally, the harmonic dispersion and DOS alone do not determine phonon lifetimes: the dense, relatively flat optical manifold indicates low optical group velocities and provides many vibrational states, but the magnitude of anharmonic scattering must be established from third-order force constants and the phonon Boltzmann equation. Modern phonon-transport analyses likewise emphasize that group velocity, lifetime, and allowed scattering phase space must be considered together.\cite{Togo2023Phonopy,Ali2024GeS2Phonons}

The corresponding GeSe$_2$ spectrum in Fig.~\ref{fig:GeSe2_phonon_dos} also contains three acoustic and 15 optical branches without significant imaginary frequencies, confirming harmonic stability of the substituted tetragonal structure. Replacing S by the heavier Se atom softens the complete spectrum: the upper frequency decreases from approximately 13.6~THz in GeS$_2$ to 10.3~THz in GeSe$_2$. This reduction is consistent with the mass dependence of lattice vibrations and with the weaker Ge--Se bonding inferred from the smaller magnitude of the Ge--Se ICOHP discussed below. The GeSe$_2$ DOS is divided by reduced-DOS windows near 4.7--5.6 and 6.2--6.8~THz, and its strongest accumulation occurs around 7.8~THz, where several optical branches are nearly flat. In contrast, GeS$_2$ exhibits a broader principal separation between the lower and upper manifolds, approximately 7.2--9.7~THz, and retains S-dominated stretching modes up to 13.6~THz.

The downward frequency shift and redistribution of optical states provide a harmonic basis for the different heat-transport responses of the two compounds, but they do not by themselves establish which material has the shorter phonon lifetime. The lower characteristic frequencies in GeSe$_2$ increase thermal occupation at a given temperature, whereas the calculated conductivity additionally depends on branch-resolved velocities, anharmonic matrix elements, and scattering phase space. Accordingly, the dispersion/DOS comparison should be interpreted together with the calculated third-order force constants and $\kappa_l$, rather than as a standalone predictor of their conductivity ordering.\cite{Li2014ShengBTE,Togo2023Implementation}

\subsection{Spectral lattice thermal transport}

\begin{figure}[t]
\centering
\includegraphics[width=\columnwidth]{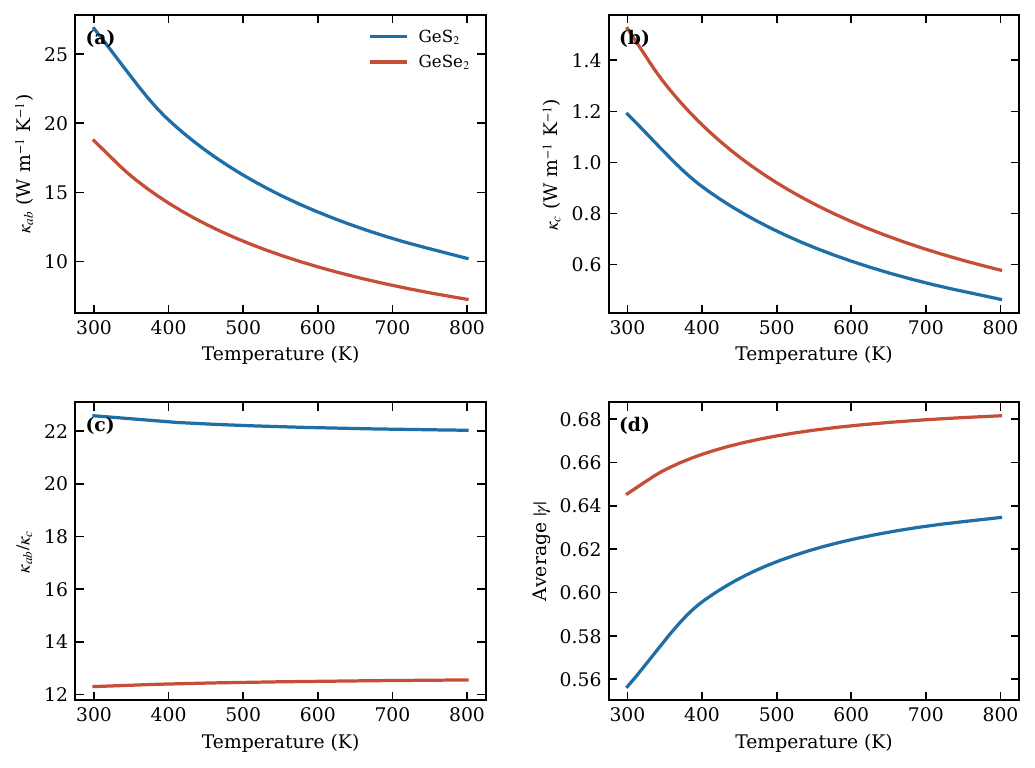}
\caption{Comparison of ShengBTE RTA lattice-transport descriptors for tetragonal GeS$_2$ and GeSe$_2$: (a) in-plane lattice thermal conductivity, (b) cross-plane lattice thermal conductivity, (c) anisotropy ratio $\kappa_{ab}/\kappa_c$, and (d) average mode Gr\"uneisen parameter.}
\label{fig:kappa_comparison}
\end{figure}

Figure~\ref{fig:kappa_comparison} summarizes the lattice-transport quantities most directly responsible for the thermal denominator of $zT$.\cite{Li2014ShengBTE} In the RTA treatment, GeS$_2$ has $\kappa_{ab}=26.86$~W~m$^{-1}$~K$^{-1}$ and $\kappa_c=1.19$~W~m$^{-1}$~K$^{-1}$ at 300~K; these values decrease to 10.22 and 0.46~W~m$^{-1}$~K$^{-1}$ at 800~K. GeSe$_2$ shows smaller in-plane but slightly larger cross-plane thermal conductivity, with $\kappa_{ab}=18.74$~W~m$^{-1}$~K$^{-1}$ and $\kappa_c=1.52$~W~m$^{-1}$~K$^{-1}$ at 300~K, decreasing to 7.25 and 0.58~W~m$^{-1}$~K$^{-1}$ at 800~K. The iterative solution gives slightly larger values but preserves the same anisotropic contrast, as shown in the Supplementary Information. Because the thermoelectric analysis below uses the RTA tensor, the RTA values are used consistently in both the phonon and $zT$ discussions. Following earlier lattice-transport analyses of low-$\kappa_l$ thermoelectrics, we interpret these values using a combined descriptor set rather than a single metric: bonding strength, Gr\"uneisen response, phonon velocities, lifetimes, and the frequency range over which heat-carrying modes accumulate all enter the final conductivity.\cite{Singh2022HeuslerLowKappa,Hossain2025ZintlXIn2C2}

The temperature dependence provides an additional check on the physical origin of the low $\kappa_c$. Recent work on crystalline materials with intrinsically ultralow, glass-like heat transport shows that weak bonding, flat low-frequency rattling modes, strong multi-phonon scattering, and wave-like coherence can produce nearly temperature-independent $\kappa_l$.\cite{Xia2025GlassLikeThermal} The present GeX$_2$ compounds behave differently: between 300 and 800~K, the calculated RTA conductivities follow approximately $T^{-0.97}$--$T^{-0.99}$ for both in-plane and cross-plane components. Thus, tetragonal GeS$_2$ and GeSe$_2$ should be described as anisotropic low-cross-plane-conductivity crystals rather than glass-like thermal conductors within the present third-order ShengBTE treatment. Their low $\kappa_c$ mainly reflects out-of-plane suppression of heat-carrying phonon velocities and finite anharmonic scattering in the layered network.

The RTA anisotropy ratio is larger in GeS$_2$ ($\kappa_{ab}/\kappa_c\approx22$) than in GeSe$_2$ ($\approx12.3$--12.5), showing that Se substitution reduces the in-plane/cross-plane contrast even though both materials remain strongly anisotropic. The average mode Gr\"uneisen parameter increases from 0.56 to 0.63 in GeS$_2$ and from 0.65 to 0.68 in GeSe$_2$ between 300 and 800~K, indicating moderate anharmonicity that strengthens with temperature. The weak-interlayer picture reported for layered GeSe$_2$ is consistent with this anisotropic response, but our calculations also show that local Ge--Se bonds remain chemically stabilizing; therefore, the low $\kappa_c$ should be assigned to anisotropic lattice dynamics rather than to a globally weak or mechanically unstable framework.\cite{Yang2019WeakInterlayerGeSe2} Together, the low cross-plane conductivity, persistent anisotropy, and finite anharmonic response explain why $zT$ is larger along $c$ than in plane.

\begin{figure}[t]
\centering
\includegraphics[width=\columnwidth]{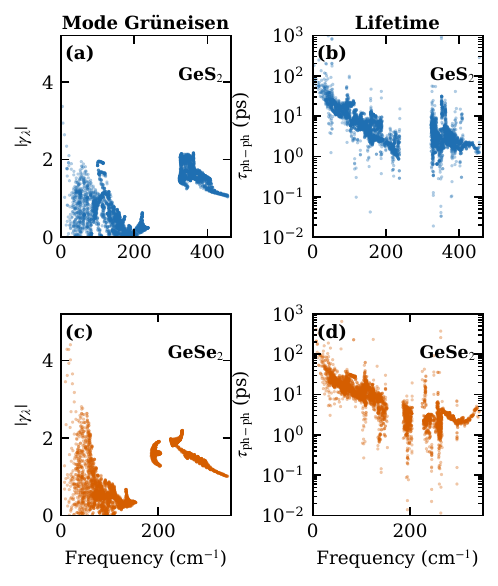}
\caption{Mode-resolved ShengBTE scattering descriptors at 300~K for tetragonal GeS$_2$ and GeSe$_2$: (a,c) absolute mode Gr\"uneisen parameters and (b,d) phonon--phonon lifetimes from final scattering rates. The cumulative conductivity, velocity, phase-space, spectral-conductivity, and iterative-tensor diagnostics are given in the Supplementary Information.}
\label{fig:spectral_scattering}
\end{figure}

Figure~\ref{fig:spectral_scattering} gives a mode-resolved view of the scattering channels behind the tensor averages in Fig.~\ref{fig:kappa_comparison}. The largest $|\gamma_\lambda|$ values occur mainly in low-frequency branches and selected optical groups, confirming that anharmonicity is distributed across several parts of the spectrum rather than confined to one isolated mode. This is important because recent first-principles thermoelectric studies show that low lattice conductivity is usually produced by the joint action of soft modes, reduced group velocities, short lifetimes, and enhanced scattering phase space, not by anharmonicity alone.\cite{Singh2022HeuslerLowKappa,Hossain2025ZintlXIn2C2} At 300~K, many acoustic and low-lying optical modes retain picosecond-to-tens-of-picoseconds phonon--phonon lifetimes, while several higher-frequency optical modes have shorter lifetimes and contribute less to heat conduction because of their smaller velocities and weaker cumulative weight. Thus the low cross-plane conductivity is not caused by a single factor; it follows from anisotropic group velocities, finite anharmonic scattering, and the limited heat carried through the stacking direction.

The cumulative curves in the Supplementary Information show that most of the RTA lattice conductivity is accumulated below relatively low frequencies. For GeS$_2$, 90\% of the in-plane cumulative conductivity is reached by about 151~cm$^{-1}$, while the cross-plane component requires modes up to about 209~cm$^{-1}$ because its total magnitude is small and more broadly distributed. For GeSe$_2$, the corresponding 90\% thresholds are lower, about 112~cm$^{-1}$ for the in-plane component and 100~cm$^{-1}$ for the cross-plane component. These lower characteristic frequencies are consistent with mass-induced softening after replacing S with Se, whereas the persistent separation between $\kappa_{ab}$ and $\kappa_c$ reflects the layered tetragonal network.

\subsection{Insight from electronic structures}

\begin{figure}[t]
\centering
\includegraphics[width=\columnwidth]{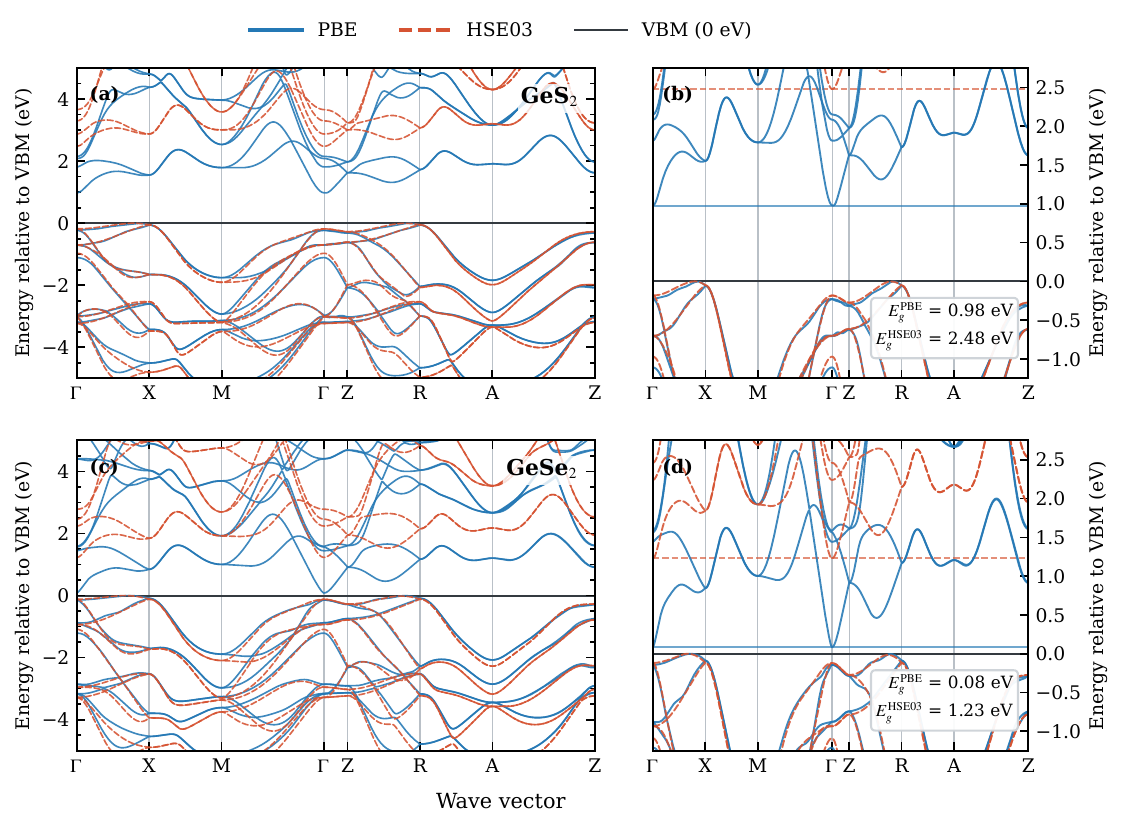}
\caption{
Wannier-interpolated PBE and HSE03 band structures of tetragonal GeS$_2$ and GeSe$_2$ along the $\Gamma$--X--M--$\Gamma$--Z--R--A--Z high-symmetry path. Panels (a) and (c) show the full dispersions, while panels (b) and (d) enlarge the corresponding band-edge regions. The y-axis in all panels is $E-E_{\mathrm{VBM}}$ in eV; panels (b) and (d) use a narrower y-axis window to resolve the band edges. For each functional, the energy zero is set to its own valence-band maximum; the colored horizontal guides and inset values identify the respective path band gaps.
}
\label{fig:band_structures}
\end{figure}

Figure~\ref{fig:band_structures} compares the PBE and HSE03 band dispersions after Wannier interpolation.\cite{Perdew1996PBE,Heyd2003HSE03,Paier2006HSEsolids,Krukau2006HSEScreening,Marzari1997MLWF,Souza2001Wannier,Yates2007WannierInterpolation,Mostofi2008Wannier90} Along the plotted high-symmetry path, the PBE gaps are 0.98~eV for GeS$_2$ and 0.08~eV for GeSe$_2$, whereas the corresponding HSE03/Wannier path gaps increase to 2.48~eV and 1.23~eV, respectively. The dense PBE mesh used for AMSET places GeSe$_2$ even closer to gap closure, giving the small overlap noted in Sec.~2; therefore, both PBE descriptions consistently indicate that semilocal GeSe$_2$ is near the semiconductor--semimetal boundary. Thus, HSE03 mainly shifts the conduction-band manifold upward while preserving the overall band topology near the valence-band edge. The much smaller gap of GeSe$_2$ reflects the stronger energetic contribution of Se-derived states near the band edges and indicates easier carrier excitation than in GeS$_2$. Because semilocal PBE underestimates semiconductor band gaps, the HSE03 results are used as the electronic-structure reference for interpreting the thermoelectric transport trends, while PAW-PBE wavefunctions are used only for the \textsc{LOBSTER} bonding analysis.

\subsection{Carrier-dependent electronic transport}

The AMSET transport tensors show a clear carrier asymmetry, as documented in the Supplementary Information. At 300~K and $|N|=10^{20}$~cm$^{-3}$, trace-averaged $\sigma$ is $4.96\times10^{4}$ (GeS$_2$) and $2.76\times10^{4}$~S~m$^{-1}$ (GeSe$_2$) for n-type transport, compared with $3.24\times10^{3}$ and $1.39\times10^{3}$~S~m$^{-1}$ for p-type transport. The corresponding n-type mobilities, 31.0 and 17.2~cm$^2$~V$^{-1}$~s$^{-1}$, exceed the p-type values of 2.02 and 0.49~cm$^2$~V$^{-1}$~s$^{-1}$. This difference combines band velocities with state-dependent scattering and cannot be attributed to effective mass or lifetime alone. Increasing concentration raises $\sigma$ sublinearly because impurity scattering strongly reduces mobility; for n-type GeS$_2$, increasing $|N|$ from $10^{20}$ to $10^{22}$~cm$^{-3}$ raises $\sigma$ by only 2.4 times while $\mu$ falls from 31.0 to 0.74~cm$^2$~V$^{-1}$~s$^{-1}$.\cite{Ganose2021AMSET,Claes2025PhononTransport}

The $\kappa_e$ curves broadly follow $\sigma$, as both sample the same conducting states. At $10^{20}$~cm$^{-3}$ and 300~K, trace-averaged n-type $\kappa_e$ is 0.313 and 0.198~W~m$^{-1}$~K$^{-1}$ for GeS$_2$ and GeSe$_2$, respectively. Heating lowers the n-type mobility in both systems, consistent with the shorter ADP+POP lifetimes at elevated temperature.\cite{Zhou2026TransportReview} The p-type response of GeSe$_2$ remains sensitive to the small PBE band overlap and should be regarded as qualitative until a gap-corrected AMSET calculation is performed. The carrier-concentration curves in the Supplementary Information connect these trends to the chemical-potential windows sampled by each carrier population. IMP scattering is included in the transport tensors but not in the plotted electron--phonon lifetime.

\begin{table}[H]
\centering
\caption{\textsc{LOBSTER} bonding descriptors for tetragonal GeS$_2$ and GeSe$_2$ obtained from PAW-PBE single-point calculations on the relaxed structures. ICOHP and ICOOP values are reported per nearest-neighbor Ge--X bond at the Fermi level.}
\label{tab:lobster_bonding}
\footnotesize
\setlength{\tabcolsep}{3pt}
\begin{tabular}{lcc}
\hline
Descriptor & GeS$_2$ & GeSe$_2$ \\
\hline
Charge spilling & 0.90\% & 0.84\% \\
Ge--X bond length (\AA{}) & 2.263 & 2.405 \\
ICOHP (eV bond$^{-1}$) & $-5.27$ & $-4.74$ \\
ICOOP & 0.286 & 0.264 \\
ICOBI & 0.900 & -- \\
Mulliken charge Ge/X ($e$) & $+0.54/-0.27$ & $+0.36/-0.18$ \\
Loewdin charge Ge/X ($e$) & $+0.43/-0.21$ & $+0.27/-0.13$ \\
\hline
\end{tabular}
\end{table}

To clarify the bonding origin of the transport response, we performed \textsc{LOBSTER} analyses of GeS$_2$ and GeSe$_2$ using PAW-PBE single-point calculations on the same relaxed structures used in the main calculations.\cite{Dronskowski1993COHP,Deringer2011COHP,Maintz2016LOBSTER,Nelson2020LOBSTER} PAW datasets were used only for this bonding step because LOBSTER reconstructs local-orbital populations and COHP/COOP/COBI descriptors from PAW-compatible wavefunctions. The PAW and ultrasoft calculations give nearly identical PBE band gaps for GeS$_2$ (about 0.98--0.99~eV), indicating that the PAW calculation does not change the qualitative electronic structure. The low absolute charge spilling values of 0.90\% for GeS$_2$ and 0.84\% for GeSe$_2$ further confirm that the projection onto local orbitals is reliable for bonding analysis.\cite{SanchezPortal1995Spilling}

The eight nearest-neighbor Ge--X bonds are nearly equivalent in both compounds, with bond lengths of 2.263~\AA{} for Ge--S and 2.405~\AA{} for Ge--Se. The ICOHP values are negative for both bonds, showing stabilizing Ge--chalcogen interactions; however, the magnitude decreases from about $-5.27$~eV per Ge--S bond to $-4.74$~eV per Ge--Se bond. The positive ICOOP values, 0.286 for GeS$_2$ and 0.264 for GeSe$_2$, indicate covalent orbital overlap in both compounds, while the Mulliken and Loewdin charges show charge transfer from Ge to the chalcogen atoms. Thus, the local Ge--S and Ge--Se bonds are best described as mixed covalent--ionic rather than purely ionic, with GeS$_2$ showing slightly stronger local covalent bonding. Because the tetragonal $P4_{2}/nmc$ structure is centrosymmetric, these local bond polarities cancel at the unit-cell level; therefore, bulk tetragonal GeS$_2$ and GeSe$_2$ are non-polar crystals despite having polar Ge--X bonds.

This bonding picture helps explain the thermoelectric behavior. Strong local Ge--X bonds preserve structural stability, but the layered tetrahedral framework remains elastically anisotropic and soft against shear, as reflected by the small $C_{66}$ values and low Debye temperatures discussed above. Such a combination of mixed covalent--ionic bonding and soft anisotropic deformation channels can suppress phonon velocities and enhance phonon scattering without destroying the semiconducting electronic structure. The LOBSTER results therefore support the view that low lattice thermal conductivity in tetragonal GeX$_2$ arises from structural anisotropy and shear softness rather than from weak local Ge--X bonding alone.

\subsection{Anisotropic thermoelectric figure of merit}

\begin{figure}[t]
\centering
\includegraphics[width=\columnwidth]{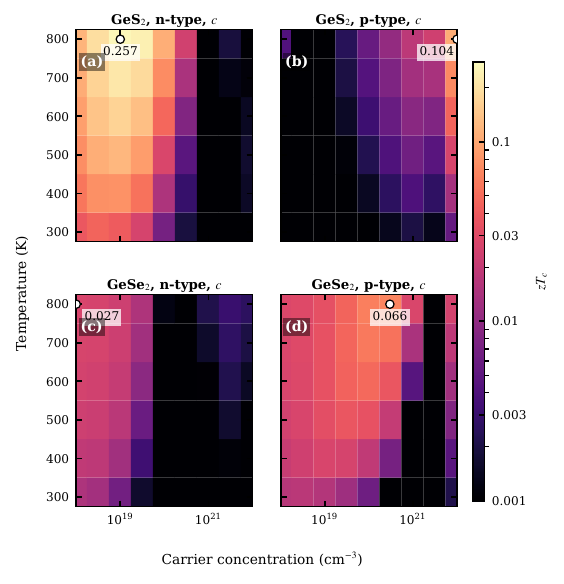}
\caption{Cross-plane figure of merit $zT_c$ of tetragonal GeS$_2$ and GeSe$_2$ obtained by combining AMSET electronic transport coefficients with the ShengBTE RTA lattice thermal conductivity component $\kappa_c$. The maps show the explicitly sampled temperature--carrier-concentration grid for n- and p-type transport; white circles mark the largest sampled value in each panel. Full component-resolved $ab$ and $c$ carrier-concentration curves are provided in the Supplementary Information.}
\label{fig:zt_summary}
\end{figure}

\begin{figure}[t]
\centering
\includegraphics[width=\columnwidth]{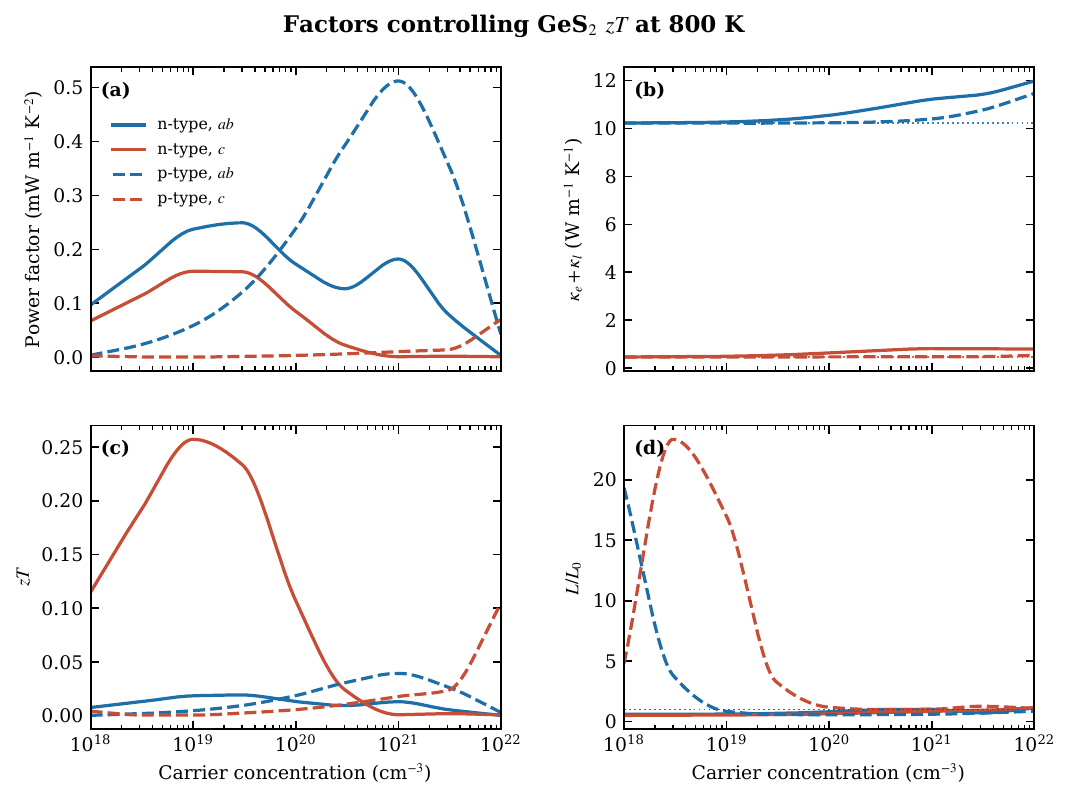}
\caption{Transport factors controlling the 800~K thermoelectric response of tetragonal GeS$_2$: (a) power factor, (b) total thermal conductivity entering the denominator of $zT$, (c) resulting $zT$, and (d) Lorenz number normalized by the Sommerfeld value $L_0=2.44\times10^{-8}$~W~$\Omega$~K$^{-2}$. The dotted horizontal guides in panel (b) mark the RTA lattice components $\kappa_{ab}$ and $\kappa_c$.}
\label{fig:GeS2_zt_factors}
\end{figure}

The anisotropic figure of merit was evaluated using Eq.~\ref{eq:zt}, with $S$, $\sigma$, and $\kappa_e$ taken from the same AMSET carrier concentration and temperature and with the corresponding component of $\kappa_l$ taken from the ShengBTE RTA tensor.\cite{Snyder2008ComplexTE,Ganose2021AMSET,Li2014ShengBTE} The tensor components were combined consistently: $xx$ electronic coefficients were paired with the in-plane lattice value $\kappa_{ab}=(\kappa_{xx}+\kappa_{yy})/2$, while the $zz$ coefficients were paired with $\kappa_c$. This component-wise reconstruction follows the same self-consistency principle emphasized in recent thermoelectric data-curation work: reported $zT$ values should be traceable to the underlying $S$, $\sigma$, $\kappa$, and temperature values rather than treated as an independent number.\cite{Ryu2025teMatDb} The heat maps in Fig.~\ref{fig:zt_summary} are plotted on the explicitly sampled carrier-concentration and temperature grid; the marked values are grid-point results rather than interpolation-derived optima. Recent first-principles analyses of thermoelectric materials emphasize the same point: $zT$ is controlled by the simultaneous balance of Seebeck response, electrical conductivity, electronic thermal conductivity, and lattice thermal conductivity, so a low $\kappa_l$ alone is not sufficient for high performance.\cite{Chaput2025ZTAbInitio,Hossain2025ZintlXIn2C2}

For n-type GeS$_2$, the calculated cross-plane value is $zT_c=0.257$ at 800~K and $n=1\times10^{19}$~cm$^{-3}$. At this point, the cross-plane power factor is only 0.159~mW~m$^{-1}$~K$^{-2}$, while $\kappa_e+\kappa_l$ is 0.496~W~m$^{-1}$~K$^{-1}$. Thus, the low $\kappa_c$ helps the denominator, but the numerator remains limited by modest cross-plane electrical conductivity ($3.0\times10^{3}$~S~m$^{-1}$) and a finite Seebeck coefficient of $-230~\mu$V~K$^{-1}$ [Fig.~\ref{fig:GeS2_zt_factors}]. At 300~K, the corresponding cross-plane value is 0.044 at $3\times10^{18}$~cm$^{-3}$. The favorable n-type concentration shifts upward with temperature because the conductivity gain from additional carriers initially outweighs the simultaneous reduction in $|S|$ and increase in $\kappa_e$. In-plane n-type performance is much smaller, reaching only $zT_{ab}=0.019$ at 800~K and $3\times10^{19}$~cm$^{-3}$, chiefly because the in-plane lattice conductivity remains large.

The p-type response follows a different concentration dependence. At 800~K, the calculated values are 0.039 in plane at $p=1\times10^{21}$~cm$^{-3}$ and 0.104 along $c$ at $p=1\times10^{22}$~cm$^{-3}$. The latter value occurs at the upper boundary of the concentration grid, so it should not be interpreted as a fully resolved optimum. The Lorenz number in Fig.~\ref{fig:GeS2_zt_factors}(d) also varies with carrier type, transport component, and concentration, indicating that a fixed Wiedemann--Franz Lorenz number would not capture the electronic heat transport quantitatively in the dilute-to-moderately doped regime.\cite{Kim2015LorenzNumber,Chaput2025ZTAbInitio} Overall, cross-plane n-type transport gives the strongest calculated $zT$, despite the larger in-plane electrical conductivity, because the more than twentyfold reduction of $\kappa_l$ along $c$ produces a substantially smaller thermal denominator. The predicted response remains moderate, indicating that pristine tetragonal GeS$_2$ is not a high-$zT$ material under the present scattering model, although its strong anisotropic contrast may be relevant when transport orientation can be controlled.

For GeSe$_2$, the updated PBE-AMSET estimate gives smaller n-type values than GeS$_2$, with $zT_c=0.027$ at 800~K and $n=1\times10^{18}$~cm$^{-3}$. For p-type cross-plane transport, the calculated value is $zT_c=0.066$ at 800~K and $p=3\times10^{20}$~cm$^{-3}$. At this point, the cross-plane power factor is 0.050~mW~m$^{-1}$~K$^{-2}$ and $\kappa_e+\kappa_l$ is 0.600~W~m$^{-1}$~K$^{-1}$, so the low cross-plane lattice contribution is partly offset by modest electrical conductivity and a finite electronic heat current. Since the GeSe$_2$ AMSET tensors are based on the semilocal PBE electronic structure, and the HSE03/Wannier calculation opens a substantially larger gap, the GeSe$_2$ $zT$ values should be regarded as qualitative until a gap-corrected scattering calculation is performed.

These $zT$ values inherit the approximations of the two transport calculations. In particular, the AMSET inputs use PBE-derived bands and approximate orbital-overlap factors, while the ShengBTE RTA treatment neglects the iterative redistribution of nonequilibrium phonon populations.\cite{Ganose2021AMSET,Li2014ShengBTE} Grain boundaries, defects, and additional high-temperature scattering channels could change $\kappa_l$, whereas gap-corrected AMSET calculations could modify the electronic coefficients, especially for GeSe$_2$.

\FloatBarrier
\section{Conclusion}

We have presented a first-principles study of tetragonal GeS$_2$ and GeSe$_2$ that connects structure, bonding, lattice dynamics, and anisotropic thermoelectric transport within one consistent framework. Both compounds are mechanically and dynamically stable in the calculated tetragonal $P4_{2}/nmc$ structure. HSE03/Wannier interpolation gives semiconducting gaps of 2.48~eV for GeS$_2$ and 1.23~eV for GeSe$_2$, while LOBSTER analysis shows mixed covalent--ionic Ge--X bonding. The stronger Ge--S interaction, reflected by the larger ICOHP magnitude ($-5.27$~eV per bond compared with $-4.74$~eV for Ge--Se), is consistent with the higher GeS$_2$ phonon-frequency range.

The central result is the strong anisotropic separation between in-plane and cross-plane heat flow. At 300~K, the ShengBTE RTA lattice thermal conductivity changes from 26.86 to 1.19~W~m$^{-1}$~K$^{-1}$ between the in-plane and $c$ components for GeS$_2$, and from 18.74 to 1.52~W~m$^{-1}$~K$^{-1}$ for GeSe$_2$. At 800~K, the corresponding cross-plane values fall to 0.46 and 0.58~W~m$^{-1}$~K$^{-1}$. This suppressed out-of-plane thermal transport arises from the layered tetrahedral framework, shear softness, low-frequency heat-carrying phonons, and moderate anharmonicity.

When these lattice tensors are combined with scattering-aware AMSET electronic transport, the calculated thermoelectric response is strongest along the cross-plane direction. The calculated value is $zT_c=0.257$ for n-type GeS$_2$ at 800~K and $n=1\times10^{19}$~cm$^{-3}$. For GeSe$_2$, the PBE-AMSET estimate gives $zT_c=0.066$ for p-type transport at 800~K and $p=3\times10^{20}$~cm$^{-3}$, although this value should be treated as qualitative because the PBE transport bands are close to a band-overlap limit. Overall, the study shows that low $\kappa_l$ alone is not enough to produce high $zT$: the power factor, mobility degradation at high carrier concentration, electronic heat transport, and Lorenz-number variation all control the final performance. Tetragonal GeX$_2$ compounds therefore emerge as anisotropic thermoelectrics with moderate calculated $zT$ values, with the most promising transport channel along the $c$ axis. Future gap-corrected scattering calculations, defect engineering, alloying, and strain control should be the most direct routes for testing whether the favorable cross-plane thermal response can be converted into higher thermoelectric efficiency.

\section*{Declaration of competing interest}

The authors declare that they have no known conflicts of interest associated with this article.

\section*{Computing facilities}

The computations were performed using the CINECA Leonardo supercomputing facility during the Master in High Performance Computing (MHPC) thesis work at the Abdus Salam International Centre for Theoretical Physics (ICTP).

\section*{CRediT authorship contribution statement}

\textbf{Neeraj Kulhari:} Writing--original draft, Validation, Methodology, Investigation, Formal analysis, Data curation.\\
\textbf{Krishna Swaroop Sharma:} Writing--review and editing, Validation, Supervision, Project administration, Conceptualization.\\
\textbf{K. C. Bhamu:} Writing--review and editing, Validation, Supervision, Project administration, Conceptualization.

\bibliographystyle{unsrtnat}
\FloatBarrier
\bibliography{references}

\clearpage
\newgeometry{margin=0.7in}
\onecolumn
\pagestyle{plain}
\fancyfoot{}
\setcounter{figure}{0}
\setcounter{table}{0}
\renewcommand{\thefigure}{S\arabic{figure}}
\renewcommand{\thetable}{S\arabic{table}}
\renewcommand{\figurename}{Fig.}
\renewcommand{\tablename}{Table}
\setlength{\parskip}{3pt}
\setlength{\parindent}{0pt}
\setlength{\floatsep}{5pt plus 1pt minus 1pt}
\setlength{\textfloatsep}{6pt plus 1pt minus 1pt}
\setlength{\intextsep}{5pt plus 1pt minus 1pt}
\captionsetup{font=small,labelfont=bf,labelsep=period,skip=3pt}
\makeatletter
\setlength{\@fptop}{0pt}
\setlength{\@fpsep}{10pt plus 2pt minus 2pt}
\setlength{\@fpbot}{0pt plus 1fil}
\makeatother

\begin{center}
{\LARGE\bfseries Supplementary Information\par}
\vspace{4pt}
{\Large Anisotropic Phonon Heat Flow and Thermoelectric Response in
Tetragonal GeS$_2$ and GeSe$_2$\par}
\vspace{14pt}
{Neeraj Kulhari$^{1}$, Krishna Swaroop Sharma$^{1}$, K. C. Bhamu$^{2,*}$\par}
\vspace{3pt}
{\small $^{1}$Department of Physics, IIS (Deemed to be University), Jaipur, Rajasthan 302020, India\par}
{\small $^{2}$Department of Physics, SLAS, Mody University of Science and Technology, Lakshmangarh, Sikar, Rajasthan 332311, India\par}
{\small $^{*}$Corresponding author: \href{mailto:kcbhamu85@gmail.com}{kcbhamu85@gmail.com}\par}
\end{center}
\vspace{12pt}

\section*{S1. Computational Settings}

The main text summarizes the methods used for the structural, elastic, electronic, phonon, and transport calculations. Table~\ref{tab:si_methods} collects the key numerical settings so that the reported figures can be reproduced more easily. The HSE03/Wannier band plots are used as band-structure references, while the quantitative thermoelectric figure of merit combines AMSET electronic coefficients with the ShengBTE relaxation-time-approximation (RTA) lattice thermal conductivity tensor.

\begin{table}[H]
\centering
\caption{Main numerical settings used in the calculations.}
\label{tab:si_methods}
\small
\begin{tabular}{p{0.25\textwidth}p{0.66\textwidth}}
\toprule
Calculation & Main settings \\
\midrule
PBE structural/electronic calculations & \textsc{Quantum ESPRESSO}; PBE+DFT-D2; scalar-relativistic ultrasoft RRKJ PSLibrary pseudopotentials for the structural, elastic, band-structure, DFPT, deformation-potential, and AMSET-input calculations; $8\times8\times4$ $k$ mesh; wave-function and charge-density cutoffs of 60 and 480~Ry. \\
HSE03 band reference & Screened hybrid functional using the same ultrasoft RRKJ PSLibrary pseudopotentials, with \texttt{exx\_fraction=0.25}, \texttt{screening\_parameter=0.106}, and $6\times6\times2$ $k$ mesh. \\
Wannier interpolation & PBE and HSE03 bands interpolated using \textsc{Wannier90}; the plotted path is $\Gamma$--X--M--$\Gamma$--Z--R--A--Z. \\
Elastic constants & \textsc{thermo\_pw} finite-strain workflow with relaxed internal coordinates, \texttt{frozen\_ions = .false.}. GeS$_2$ used six strain geometries, strain amplitude 0.005, and third-order polynomial fitting. GeSe$_2$ used four strain geometries, strain amplitude 0.002, and second-order polynomial fitting to stabilize the strained-structure relaxations. \\
AMSET transport & ADP, POP, and IMP scattering. AMSET post-processing interpolation meshes: $81\times81\times25$ for GeS$_2$ and $81\times81\times27$ for GeSe$_2$; these are separate from the Quantum ESPRESSO SCF/NSCF $k$ meshes. Carrier concentrations span $10^{18}$ to $10^{22}$~cm$^{-3}$. \\
BoltzTraP2 diagnostics & HSE03/Wannier bands post-processed in the constant-relaxation-time approximation to inspect chemical-potential trends in $S$, $\sigma/\tau$, $\kappa_e/\tau$, power factor, Lorenz number, and DOS. \\
Phonons and ShengBTE & Harmonic force constants from DFPT using the same PBE+DFT-D2 ultrasoft pseudopotential workflow; non-analytical correction from Born effective charges and dielectric tensor; third-order force constants from $4\times4\times2$ finite-displacement supercells generated with \texttt{thirdorder\_espresso.py}; \texttt{thirdorder\_espresso.py scf.in reap 4 4 2 -7} used for both compounds to retain interactions through the seventh-nearest-neighbor shell; ShengBTE transport sampled on a $20\times20\times6$ phonon mesh. \\
\bottomrule
\end{tabular}
\end{table}

\section*{S2. Electronic-Structure and BoltzTraP2 Diagnostics}

\begin{figure}[!htbp]
\centering
\includegraphics[width=0.95\textwidth]{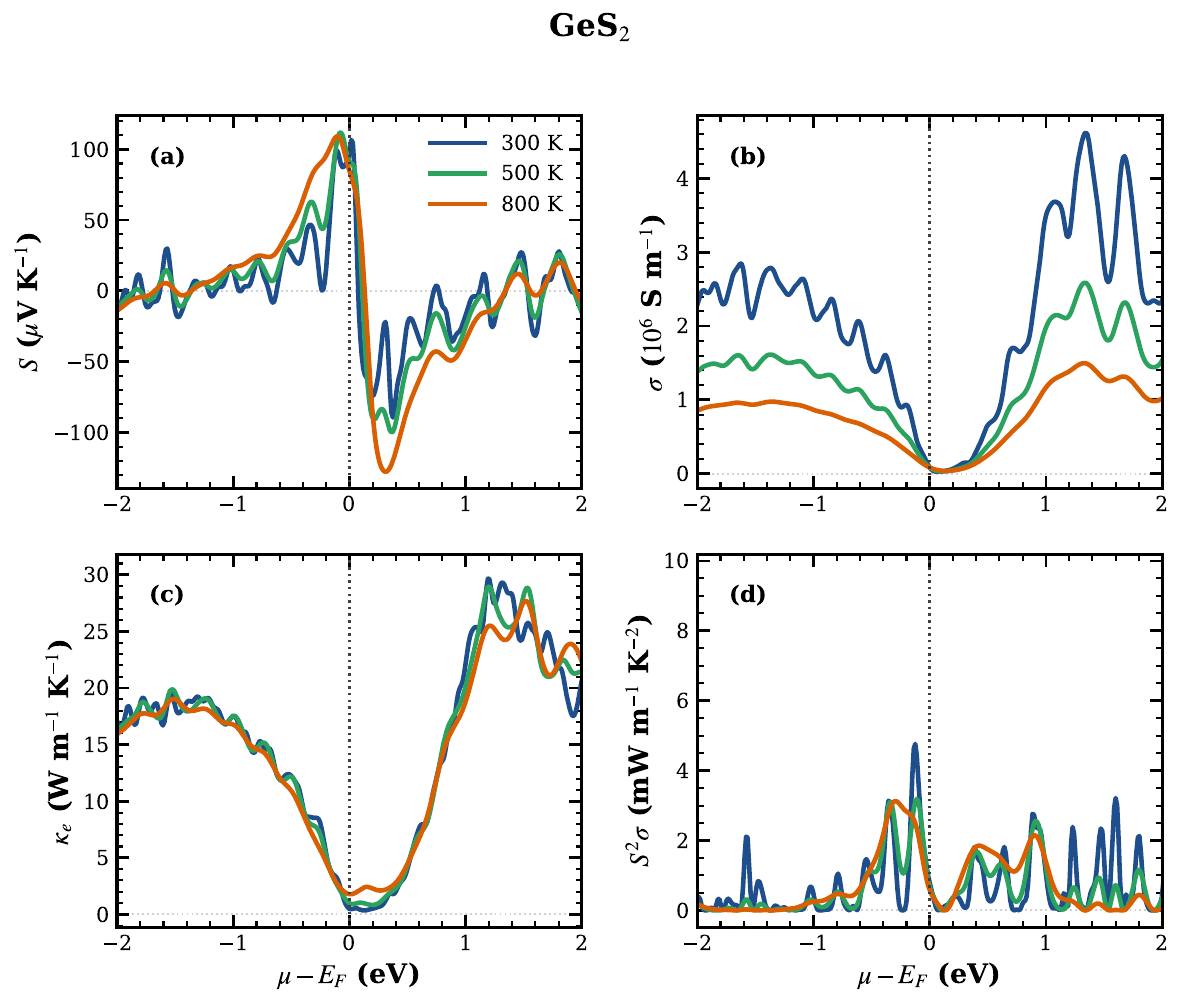}
\caption{HSE03/Wannier BoltzTraP2 transport coefficients of GeS$_2$ as functions of chemical potential for 300, 500, and 800~K.}
\label{fig:si_btp2_abs_ges2}
\end{figure}

\begin{figure}[!htbp]
\centering
\includegraphics[width=0.95\textwidth]{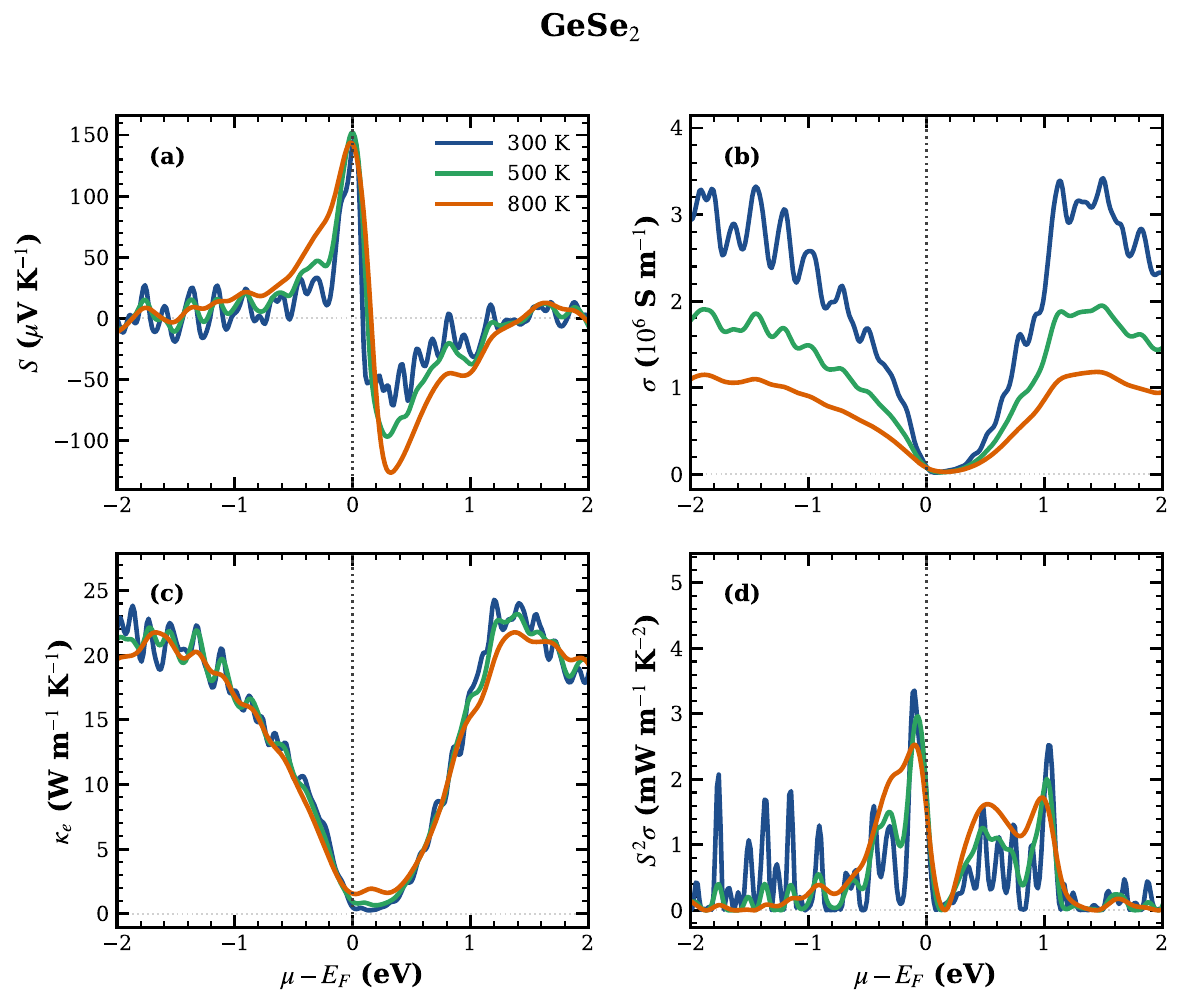}
\caption{HSE03/Wannier BoltzTraP2 transport coefficients of GeSe$_2$ as functions of chemical potential for 300, 500, and 800~K.}
\label{fig:si_btp2_abs_gese2}
\end{figure}

Figures~\ref{fig:si_btp2_abs_ges2} and \ref{fig:si_btp2_abs_gese2} show the HSE03/Wannier BoltzTraP2 transport curves plotted against chemical potential relative to the Fermi level. These plots are useful for identifying the carrier-type asymmetry and the sensitivity of $S$, $\sigma/\tau$, $\kappa_e/\tau$, and PF to the band-edge shape. They are not used directly for the final $zT$ values because the main thermoelectric estimates use AMSET state-dependent scattering.

\begin{figure}[!htbp]
\centering
\begin{subfigure}[t]{0.95\textwidth}
    \centering
    \includegraphics[width=\linewidth]{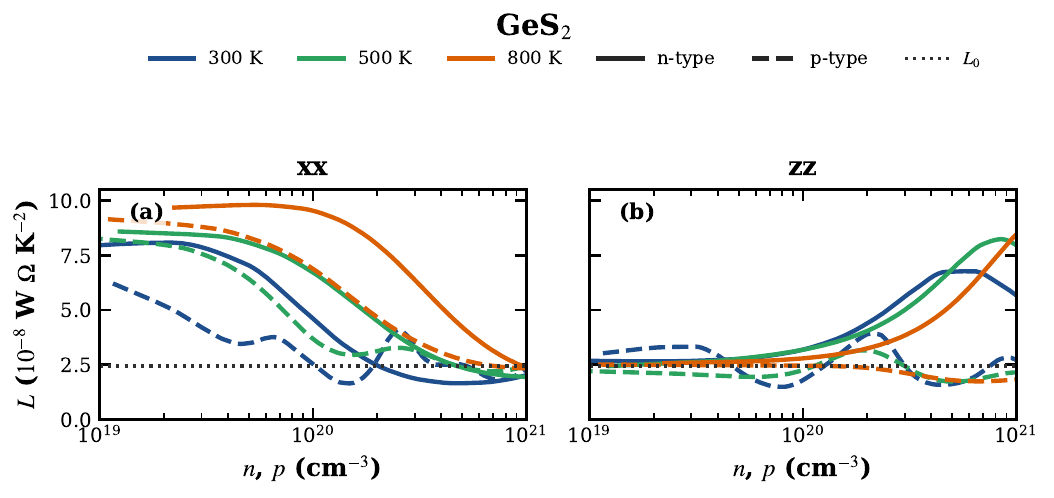}
    \caption{GeS$_2$}
\end{subfigure}
\par\vspace{3pt}
\begin{subfigure}[t]{0.95\textwidth}
    \centering
    \includegraphics[width=\linewidth]{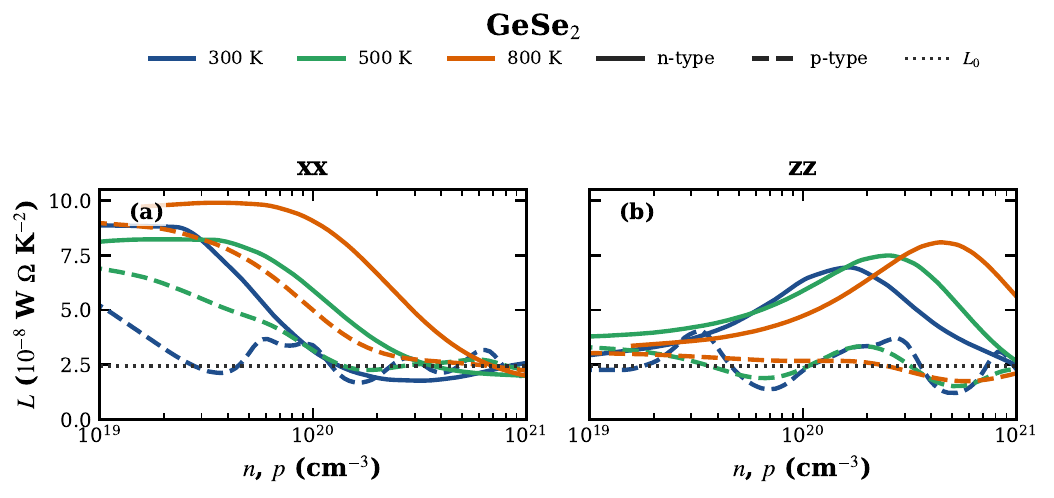}
    \caption{GeSe$_2$}
\end{subfigure}
\caption{Lorenz number obtained from the BoltzTraP2 HSE03/Wannier transport tensors.}
\label{fig:si_lorenz}
\end{figure}

The Lorenz number, $L=\kappa_e/(\sigma T)$, is shown in Fig.~\ref{fig:si_lorenz}. A constant Wiedemann--Franz value is a good approximation only for highly degenerate carriers. The present curves vary with chemical potential, temperature, and direction, showing that a fixed Lorenz number would be an oversimplification in the dilute and near-band-edge regions.

\begin{figure}[!htbp]
\centering
\includegraphics[width=0.95\textwidth]{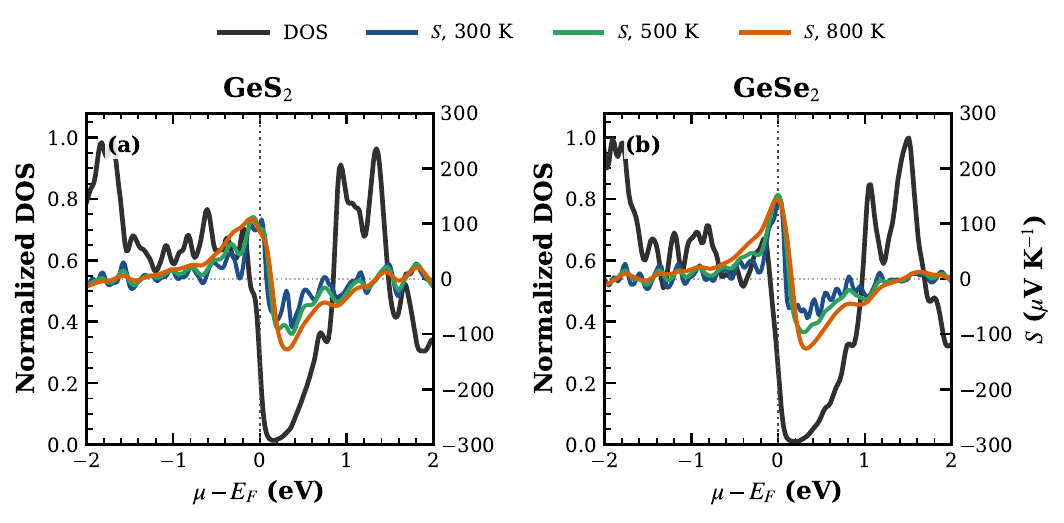}
\caption{Comparison of electronic DOS and Seebeck coefficient trends from the HSE03/Wannier BoltzTraP2 calculations.}
\label{fig:si_dos_seebeck}
\end{figure}

The electronic DOS from the trace files is plotted together with the Seebeck coefficient in Fig.~\ref{fig:si_dos_seebeck}. Peaks in the DOS close to the band edges are relevant because a sharp transport distribution can enhance the Seebeck response, as expected from the Mahan--Sofo picture of thermoelectric optimization~[1].

\section*{S3. AMSET Carrier and Scattering Diagnostics}

\begin{figure}[!htbp]
\centering
\includegraphics[width=0.95\textwidth]{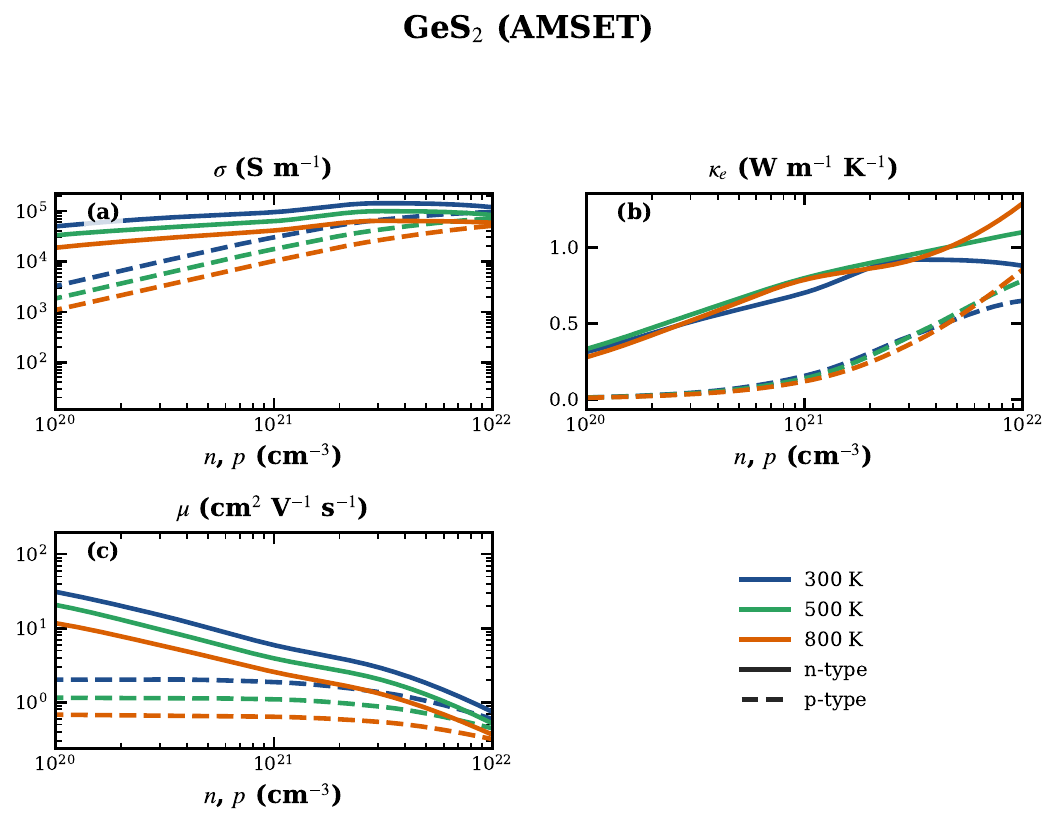}
\caption{AMSET electrical conductivity, electronic thermal conductivity, and mobility of GeS$_2$ as functions of electron and hole concentration.}
\label{fig:si_amset_conc_trace_ges2}
\end{figure}

\begin{figure}[!htbp]
\centering
\includegraphics[width=0.95\textwidth]{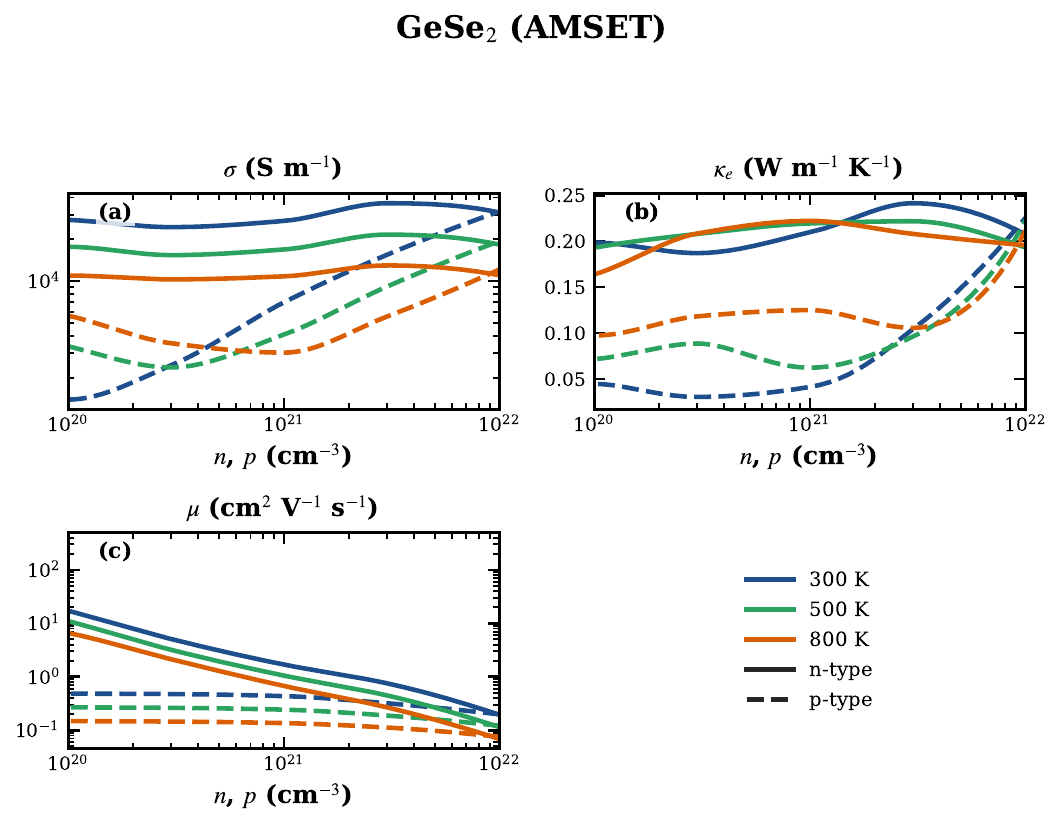}
\caption{AMSET electrical conductivity, electronic thermal conductivity, and mobility of GeSe$_2$ as functions of electron and hole concentration.}
\label{fig:si_amset_conc_trace_gese2}
\end{figure}

The AMSET carrier-dependent transport curves in Figs.~\ref{fig:si_amset_conc_trace_ges2} and \ref{fig:si_amset_conc_trace_gese2} show how $\sigma$, $\kappa_e$, and mobility vary with the sampled electron and hole concentrations. The separate 300~K concentration--chemical-potential plots below connect these carrier densities to the chemical-potential window used in the transport tensors. This is especially important for GeSe$_2$, where the PBE transport bands are close to a band-overlap limit.

\begin{figure}[!htbp]
\centering
\includegraphics[width=0.95\textwidth]{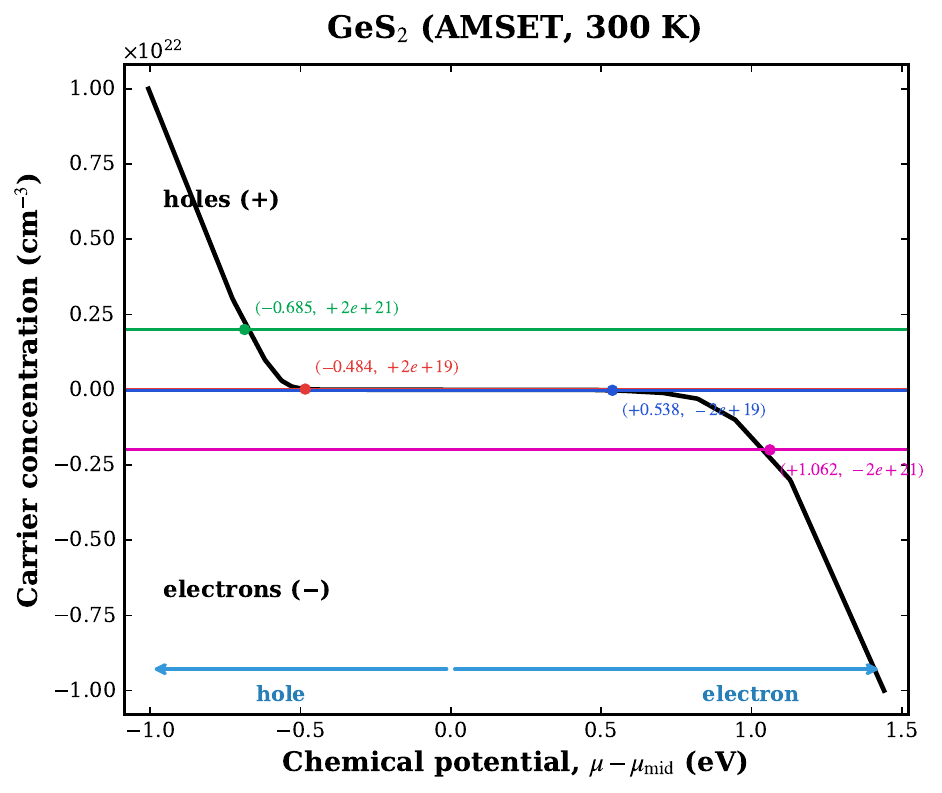}
\caption{Electron and hole concentrations of GeS$_2$ at 300~K as functions of chemical potential.}
\label{fig:si_amset_conc_300_ges2}
\end{figure}

\begin{figure}[!htbp]
\centering
\includegraphics[width=0.95\textwidth]{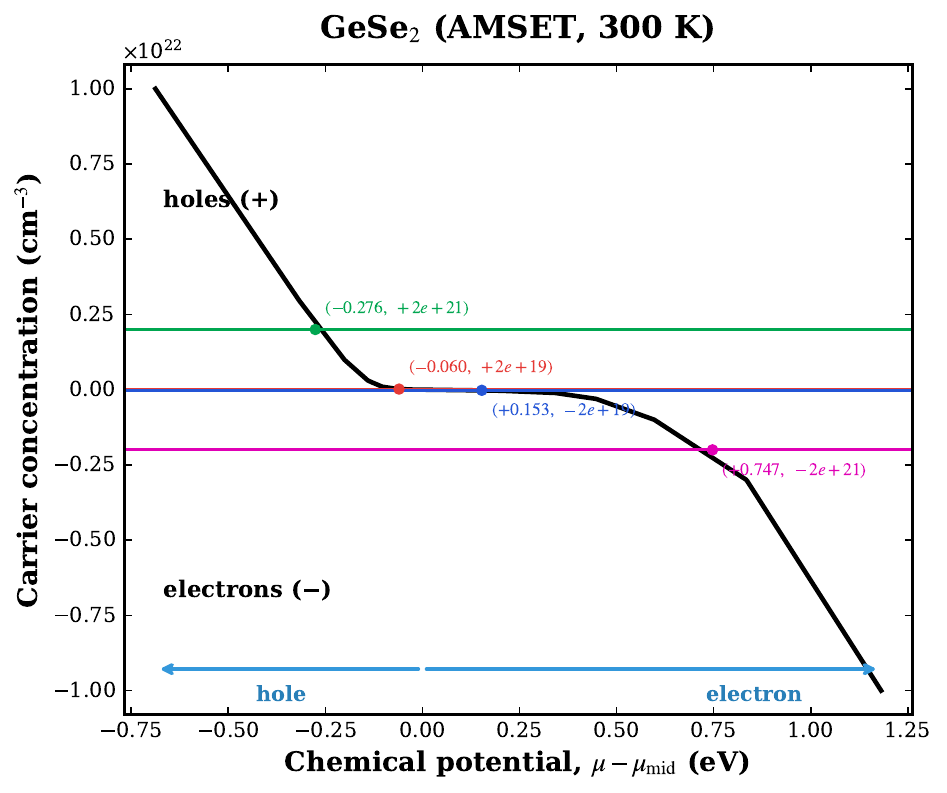}
\caption{Electron and hole concentrations of GeSe$_2$ at 300~K as functions of chemical potential.}
\label{fig:si_amset_conc_300_gese2}
\end{figure}

\begin{figure}[!htbp]
\centering
\includegraphics[width=0.95\textwidth]{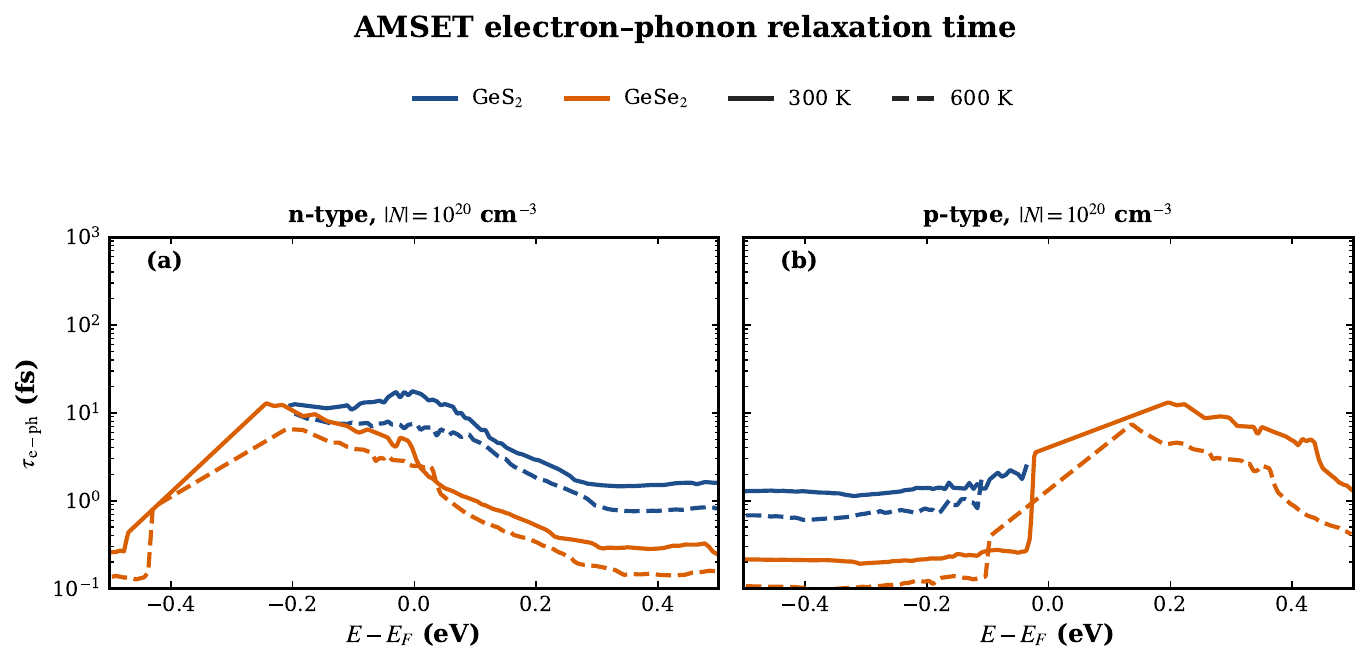}
\caption{AMSET relaxation-time comparison for GeS$_2$ and GeSe$_2$. The plotted lifetime contains the ADP and POP contributions; ionized-impurity scattering is included in the AMSET transport tensors but is not shown in this lifetime plot.}
\label{fig:si_amset_tau}
\end{figure}

\section*{S4. Anisotropic zT Data}

\begin{figure}[!htbp]
\centering
\includegraphics[width=0.95\textwidth]{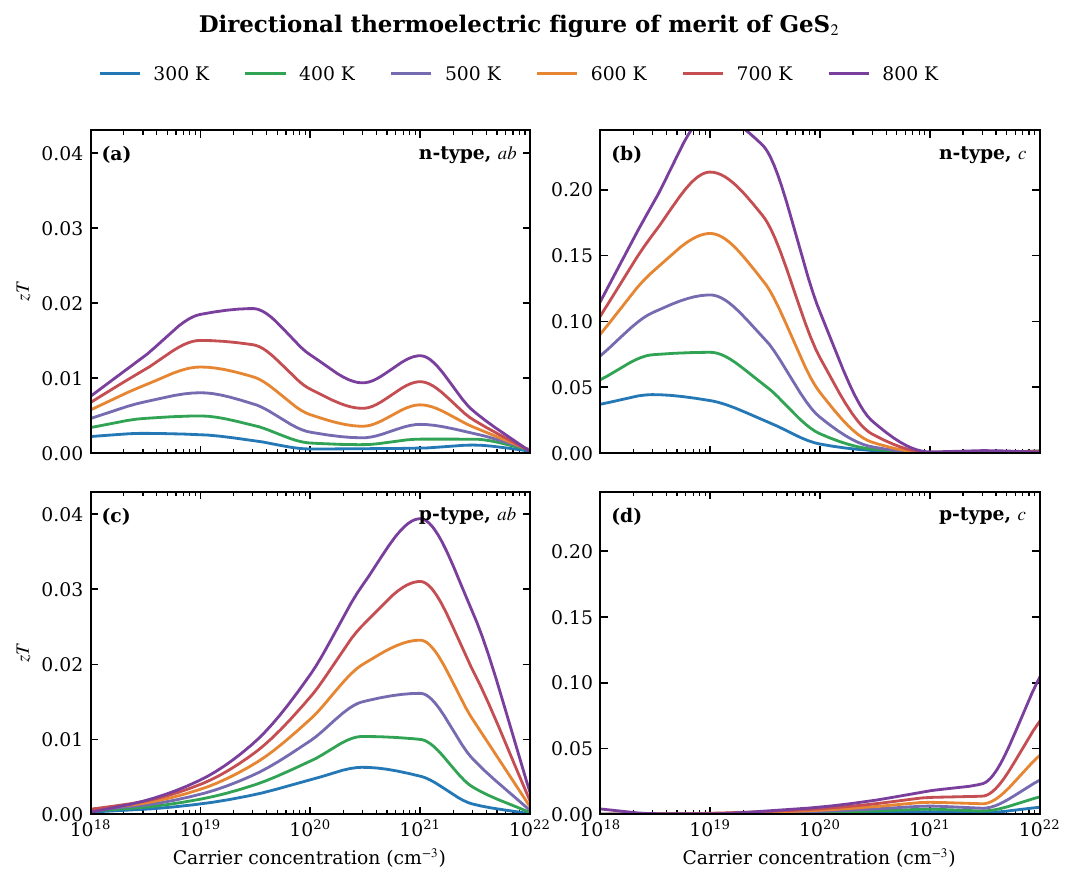}
\caption{Carrier-concentration-dependent component-resolved $zT$ values of GeS$_2$ obtained by combining AMSET electronic coefficients with ShengBTE RTA lattice thermal conductivity.}
\label{fig:si_full_zt_ges2}
\end{figure}

\begin{figure}[!htbp]
\centering
\includegraphics[width=0.95\textwidth]{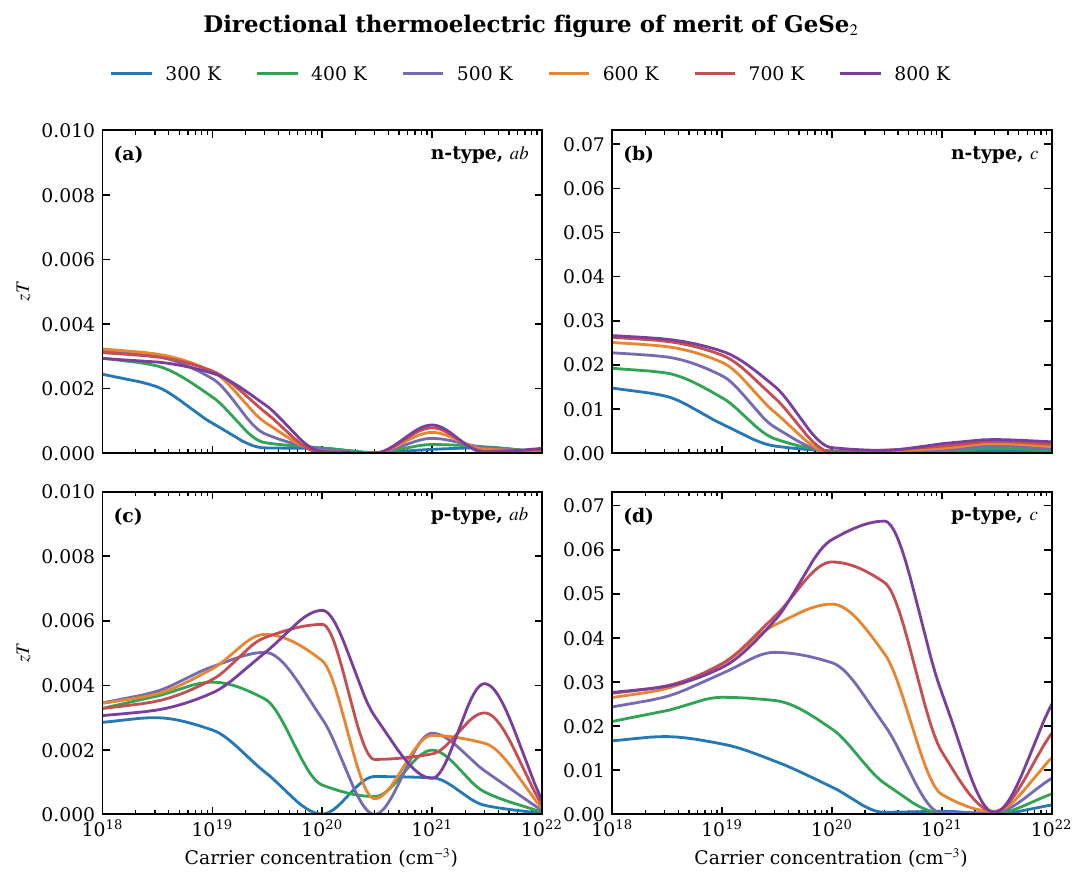}
\caption{Carrier-concentration-dependent component-resolved $zT$ values of GeSe$_2$ obtained by combining AMSET electronic coefficients with ShengBTE RTA lattice thermal conductivity.}
\label{fig:si_full_zt_gese2}
\end{figure}

The main text reports selected $zT$ values. Figures~\ref{fig:si_full_zt_ges2} and \ref{fig:si_full_zt_gese2} give the full component-resolved curves as functions of carrier concentration.

\begin{figure}[!htbp]
\centering
\includegraphics[width=0.95\textwidth]{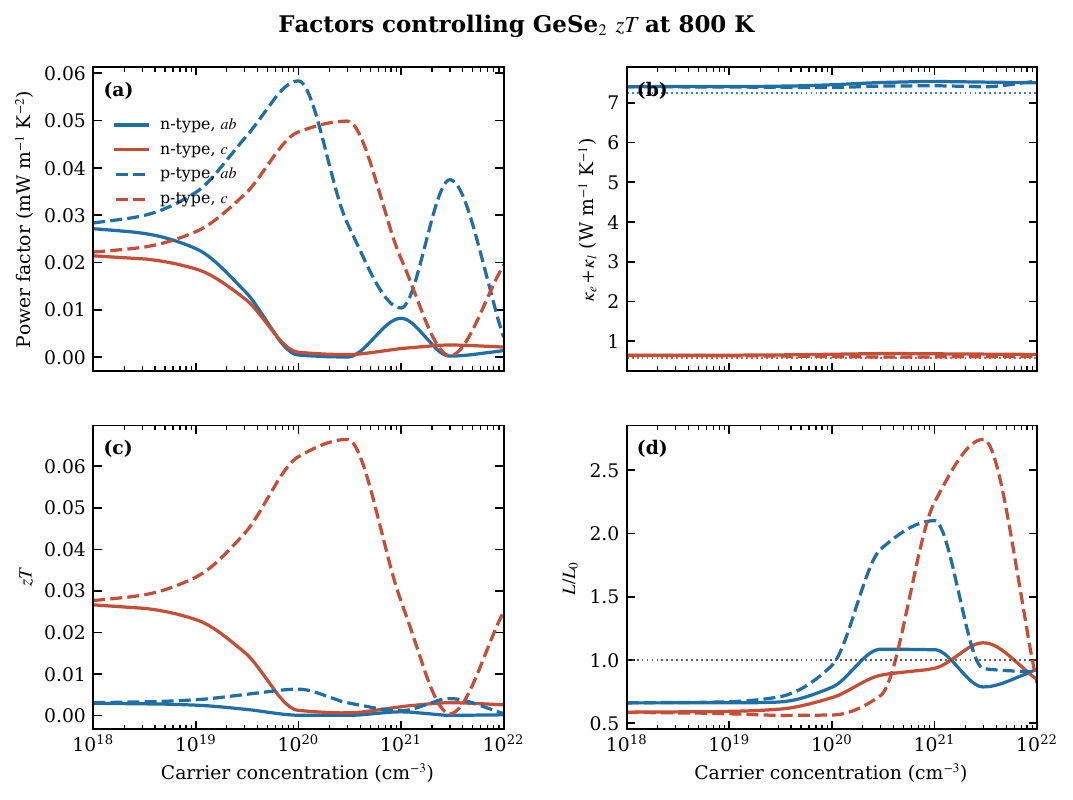}
\caption{Transport factors controlling the 800~K thermoelectric response of tetragonal GeSe$_2$: (a) power factor, (b) total thermal conductivity entering the denominator of $zT$, (c) resulting $zT$, and (d) Lorenz number normalized by the Sommerfeld value $L_0=2.44\times10^{-8}$~W~$\Omega$~K$^{-2}$. This plot is the GeSe$_2$ counterpart of the GeS$_2$ factor decomposition shown in the main text.}
\label{fig:si_gese2_zt_factors}
\end{figure}

\begin{table}[H]
\centering
\caption{Directional $zT$ values at 800~K from the sampled carrier-concentration grid. The listed concentration is the grid point where the largest value for that branch occurs, not an interpolated optimum.}
\label{tab:si_zt_800}
\small
\begin{tabular}{llcc}
\toprule
Compound & Branch & Concentration (cm$^{-3}$) & $zT$ \\
\midrule
GeS$_2$ & n-type, $ab$ & $3\times10^{19}$ & 0.019 \\
GeS$_2$ & n-type, $c$ & $1\times10^{19}$ & 0.257 \\
GeS$_2$ & p-type, $ab$ & $1\times10^{21}$ & 0.039 \\
GeS$_2$ & p-type, $c$ & $1\times10^{22}$ & 0.104 \\
GeSe$_2$ & n-type, $ab$ & $1\times10^{18}$ & 0.003 \\
GeSe$_2$ & n-type, $c$ & $1\times10^{18}$ & 0.027 \\
GeSe$_2$ & p-type, $ab$ & $1\times10^{20}$ & 0.006 \\
GeSe$_2$ & p-type, $c$ & $3\times10^{20}$ & 0.066 \\
\bottomrule
\end{tabular}
\end{table}

\section*{S5. ShengBTE Lattice-Thermal-Transport Details}

The main manuscript contains the compact mode-resolved comparison used for the discussion. The figures below retain the full per-material ShengBTE diagnostics: RTA and iterative conductivity tensors, anisotropy, cumulative conductivity, average Gr\"uneisen parameters, frequency-resolved velocities, three-phonon phase space, lifetimes, and spectral conductivity. Table~\ref{tab:si_kappa} lists the RTA values used in the main $zT$ calculation.

\begin{figure}[!htbp]
\centering
\includegraphics[width=0.95\textwidth]{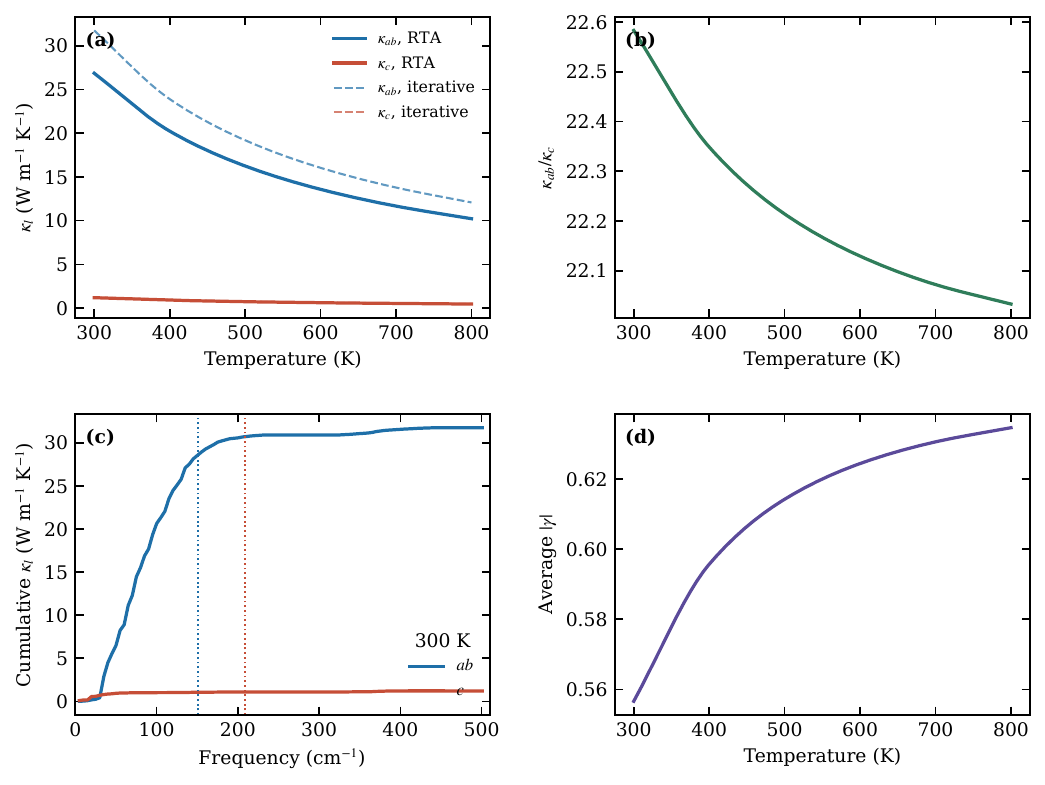}
\caption{Detailed ShengBTE lattice-transport summary for GeS$_2$.}
\label{fig:si_ges2_kappa}
\end{figure}

\begin{figure}[!htbp]
\centering
\includegraphics[width=0.95\textwidth]{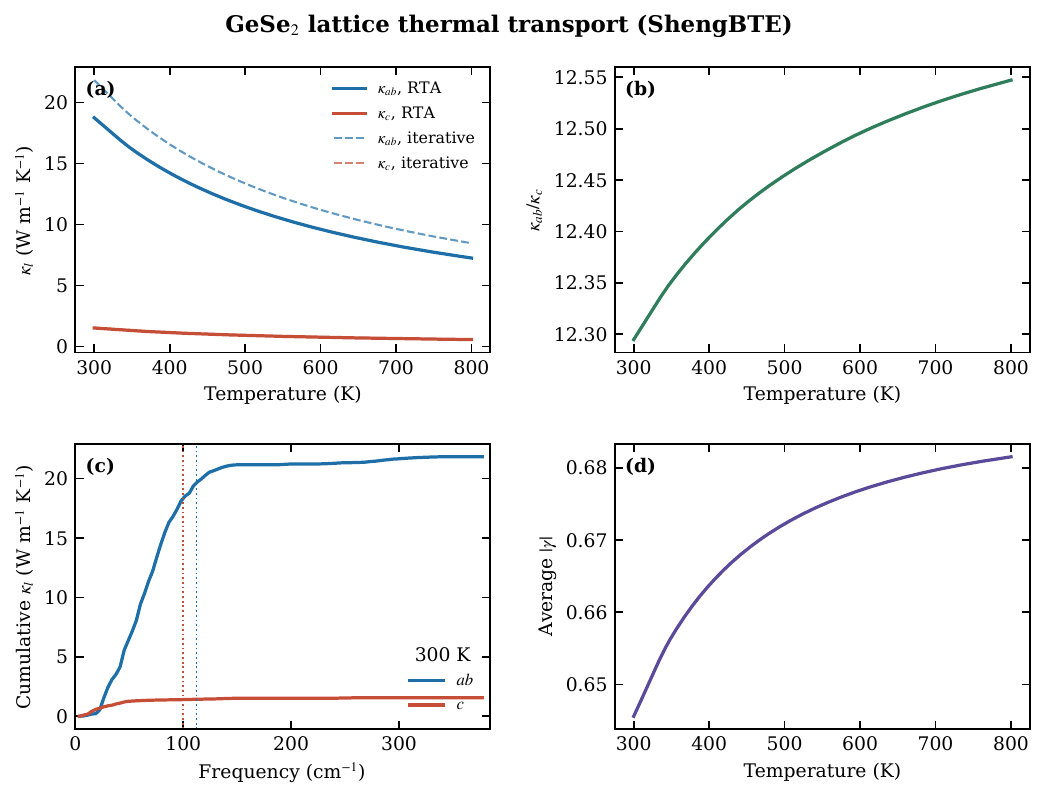}
\caption{Detailed ShengBTE lattice-transport summary for GeSe$_2$.}
\label{fig:si_gese2_kappa}
\end{figure}

\begin{figure}[!htbp]
\centering
\includegraphics[width=0.95\textwidth]{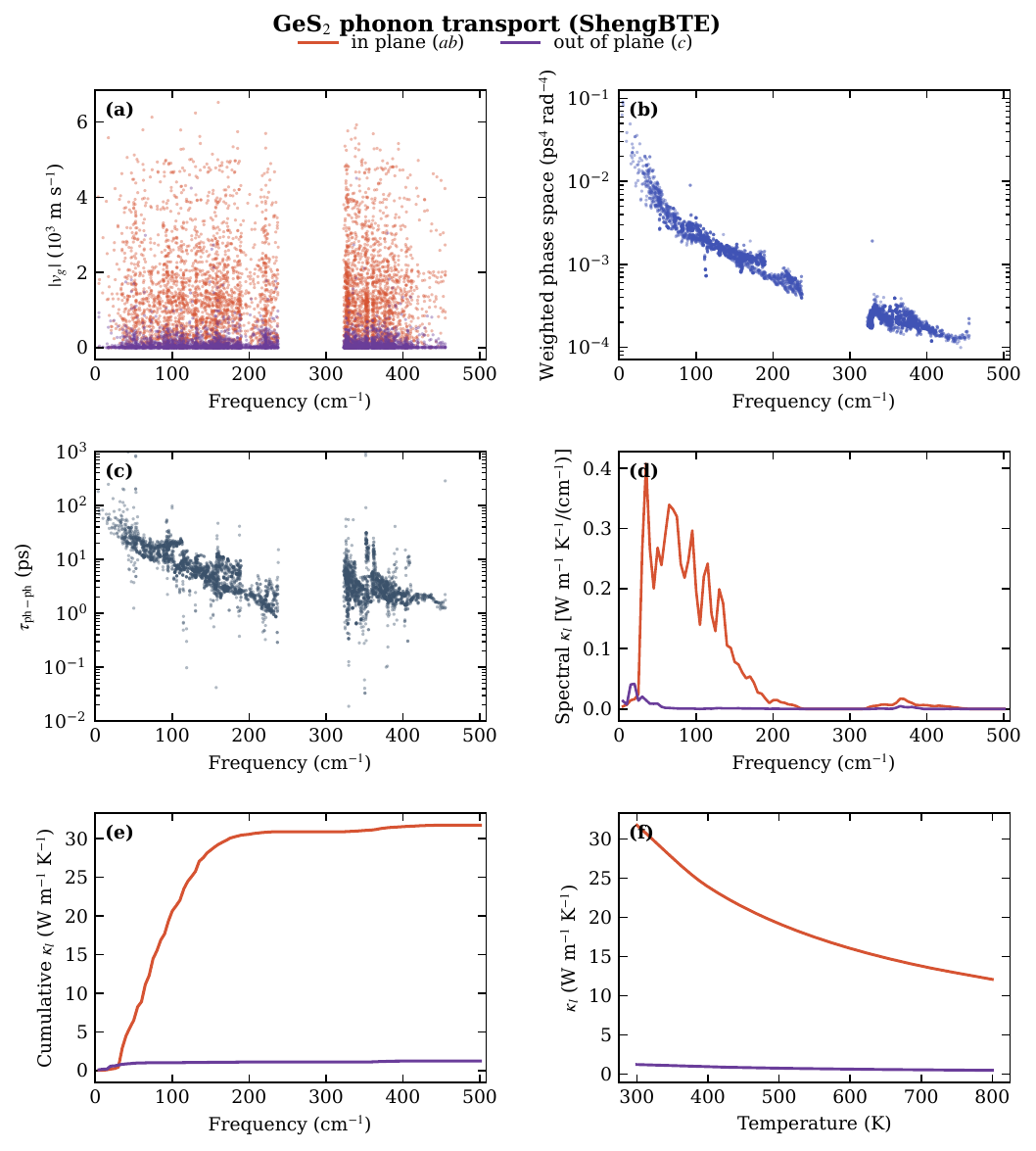}
\caption{Additional GeS$_2$ spectral phonon-transport descriptors from ShengBTE, including branch-resolved and cumulative quantities.}
\label{fig:si_ges2_spectral}
\end{figure}

\begin{figure}[!htbp]
\centering
\includegraphics[width=0.95\textwidth]{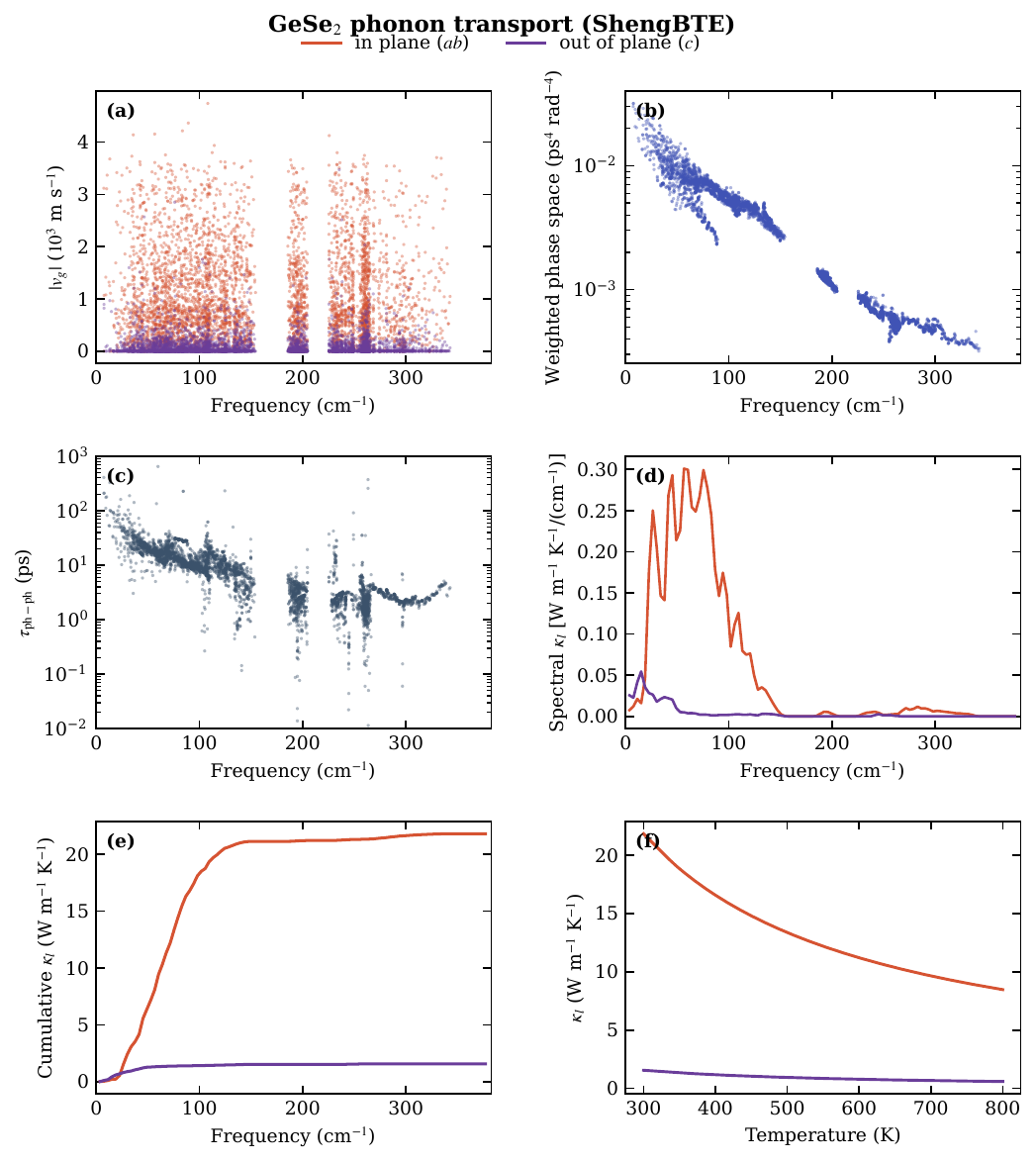}
\caption{Additional GeSe$_2$ spectral phonon-transport descriptors from ShengBTE, including branch-resolved and cumulative quantities.}
\label{fig:si_gese2_spectral}
\end{figure}

\begin{table}[H]
\centering
\caption{Selected RTA lattice thermal conductivities used in the thermoelectric analysis.}
\label{tab:si_kappa}
\small
\begin{tabular}{lcccc}
\toprule
Compound & $T$ (K) & $\kappa_{ab}$ (W m$^{-1}$ K$^{-1}$) & $\kappa_c$ (W m$^{-1}$ K$^{-1}$) & $\kappa_{ab}/\kappa_c$ \\
\midrule
GeS$_2$ & 300 & 26.86 & 1.19 & 22.58 \\
GeS$_2$ & 500 & 16.25 & 0.73 & 22.21 \\
GeS$_2$ & 800 & 10.22 & 0.46 & 22.03 \\
GeSe$_2$ & 300 & 18.74 & 1.52 & 12.29 \\
GeSe$_2$ & 500 & 11.46 & 0.92 & 12.45 \\
GeSe$_2$ & 800 & 7.25 & 0.58 & 12.55 \\
\bottomrule
\end{tabular}
\end{table}

\FloatBarrier
\section*{S6. LOBSTER Bonding Setup}

The \textsc{LOBSTER} projections were performed using PAW-PBE single-point calculations because LOBSTER requires PAW-compatible wavefunctions. For GeS$_2$, the local basis was Ge 3d 4s 4p and S 3s 3p. For GeSe$_2$, the basis was Ge 3d 4s 4p and Se 4s 4p. COHP, COOP, and COBI pairs were generated over nearest-neighbor Ge--X distances; the resulting charge spilling values were below 1\% for both compounds. The main manuscript reports the averaged nearest-neighbor ICOHP, ICOOP, and charge-transfer descriptors.

\FloatBarrier

\end{document}